\PassOptionsToPackage{table,dvipsnames}{xcolor}
\documentclass[sigconf,nonacm]{acmart}

\makeatletter
\def\@ACM@checkaffil{
    \if@ACM@instpresent\else
    \ClassWarningNoLine{\@classname}{No institution present for an affiliation}%
    \fi
    \if@ACM@citypresent\else
    \ClassWarningNoLine{\@classname}{No city present for an affiliation}%
    \fi
    \if@ACM@countrypresent\else
        \ClassWarningNoLine{\@classname}{No country present for an affiliation}%
    \fi
}
\makeatother

\setcopyright{none}
\renewcommand\footnotetextcopyrightpermission[1]{}
\AtBeginDocument{%
  \providecommand\BibTeX{{%
    \normalfont B\kern-0.5em{\scshape i\kern-0.25em b}\kern-0.8em\TeX}}}

\usepackage[normalem]{ulem}

\usepackage{tikz}
\usepackage{amsmath}

\usepackage{filecontents}

\usepackage[hang,flushmargin]{footmisc}
\usepackage{microtype} 
\usepackage{siunitx}
\usepackage[ruled,linesnumbered]{algorithm2e}
\usepackage[utf8]{inputenc}
\usepackage{framed}
\usepackage{multirow}
\usepackage{listings}
\usepackage{verbatim}
\usepackage{tikz}
\usepackage{appendix}
\usepackage{graphicx}
\usepackage{float}
\usepackage{subfig}
\usepackage{environ}
\usepackage{comment}
\usepackage{booktabs}
\usepackage{ragged2e}
\usepackage{lipsum}
\usepackage{afterpage}
\usepackage{tabularx}
\usepackage{calc}
\usepackage{flushend}
\usepackage{bm}

\usepackage{amsbsy}

\usepackage{tablefootnote}
\usepackage[para]{threeparttable}

\usepackage{pifont}%
\usepackage{listings}

\definecolor{charcoal}{rgb}{0.21, 0.27, 0.31}
\definecolor{Gray}{rgb}{0.5, 0.5, 0.5}
\definecolor{slateGray}{rgb}{0.44, 0.5, 0.56}
\definecolor{smoke}{rgb}{0.52, 0.53, 0.52}
\definecolor{steelGray}{rgb}{0.44, 0.47, 0.49}

\definecolor{grey}{rgb}{0.02, 0.02, 0.01}

\newcommand{\cmark}{\textcolor{JungleGreen}{\ding{51}}}%
\newcommand{\xmark}{\textcolor{Maroon}{\ding{55}}}%

\usepackage[varqu]{zi4}%
\AtBeginDocument{%
}

\definecolor{Maroon}{cmyk}{0, 0.87, 0.68, 0.32}
\definecolor{darkgreen}{rgb}{0.0, 0.2, 0.13}
\definecolor{JungleGreen}{rgb}{0.16, 0.67, 0.53}

\newcommand\addvmargin[1]{
  \node[fit=(current bounding box),inner ysep=#1,inner xsep=0]{};
}
\usetikzlibrary{fit}

\newcommand*\emptycirc[1][1ex]{%
  \begin{tikzpicture}[baseline=-4]
  \draw (0,0) circle (#1);
  \addvmargin{1mm}
  \end{tikzpicture}}%

\newcommand*\halfcirc[1][1ex]{%
  \begin{tikzpicture}[baseline=-4]
  \draw[fill] (0,0)-- (90:#1) arc (90:270:#1) -- cycle ;
  \draw (0,0) circle (#1);
  \addvmargin{1mm}
  \end{tikzpicture}}%

\newcommand*\fullcirc[1][1ex]{%
  \begin{tikzpicture}[baseline=-4]
  \fill (0,0) circle (#1);
  \addvmargin{1mm}
  \end{tikzpicture}}%

\definecolor{codegreen}{rgb}{0,0.56,0.56}
\definecolor{codegray}{rgb}{0.3,0.3,0.3}
\definecolor{codepurple}{rgb}{0.7,0,0.82}
\definecolor{codeblack}{rgb}{0,0,0}
\definecolor{backcolor}{rgb}{0.99,0.99,0.99}
\definecolor{keywordblue}{rgb}{0.1,0.1,1}
\definecolor{commentgray}{rgb}{0.65,0.65,0.65}

\lstdefinestyle{mystyle}{
    frame=single,
    backgroundcolor=\color{backcolor},   
    commentstyle=\sffamily\it\color{commentgray},
    keywordstyle=\color{keywordblue},
    numberstyle=\tiny\color{codeblack},
    stringstyle=\color{codepurple},
    basicstyle=\rmfamily\scriptsize\color{black},
    breakatwhitespace=false,         
    breaklines=true,                 
    captionpos=b,                    
    keepspaces=true,                 
    numbers=left,                    
    numbersep=4pt,                  
    showspaces=false,                
    showstringspaces=false,
    showtabs=false,                  
    tabsize=4,
    float=tp,
    floatplacement=tbp,
    abovecaptionskip=0pt
}

\definecolor{darkbluecolor}{rgb}{0.12,0.3,0.57}

\newcommand{\dcircle}[1]{{\color{darkbluecolor}\ding{\numexpr181 + #1}}}
\newcommand{\step}[1]{\dcircle{#1}}

\newcommand{\name}{\textsc{SpecControl}}

\newcommand{\resNo}{$Res_{no}$}
\newcommand{\resFE}{$Res_{FE}$}
\newcommand{\resBE}{$Res_{BE}$}
\newcommand{\bdInvalid}{$BD_{invalid}$}
\newcommand{\bdNo}{$BD_{no}$}
\newcommand{\bdValid}{$BD_{valid}$}
\newcommand{\brInvalid}{$BR_{invalid}$}
\newcommand{\brNo}{$BR_{no}$}
\newcommand{\brValid}{$BR_{valid}$}

\newcommand{\DOLMAPerfOverhead}{$51 \%$}
\newcommand{\sttPerfOverhead}{$43 \%$}
\newcommand{\SBPerfOverhead}{$2.26 \times$}
\newcommand{\SpecControlOverhead}{$23 \%$}
\newcommand{\DOLMAPerfImprov}{$2.22 \times$}
\newcommand{\STTPerfImprov}{$1.87 \times$}

\newcommand{\xalancSpecControlOverhead}{$18 \%$}
\newcommand{\xalancDOLMASTTOverhead}{$110 \%$}

\begin{document}

\title{Mitigating Speculation-based Attacks through Configurable Hardware/Software Co-design} 

\author{Ali Hajiabadi}
\affiliation{%
  \institution{National University of Singapore}
}

\author{Archit Agarwal}
\affiliation{%
  \institution{National University of Singapore}
}

\author{Andreas Diavastos}
\affiliation{%
  \institution{National University of Singapore}
}

\author{Trevor E. Carlson}
\affiliation{%
  \institution{National University of Singapore}
}

\begin{abstract}

New speculation-based attacks that affect large numbers of modern systems are disclosed regularly. Currently, CPU vendors regularly fall back to heavy-handed mitigations like using barriers or enforcing strict programming guidelines resulting in significant performance overhead. What is missing is a solution that allows for efficient mitigation and is flexible enough to address both current and future speculation vulnerabilities, without additional hardware changes.

In this work, we present \name{}, a novel hardware/software co-design, that enables new levels of security while reducing the performance overhead that has been demonstrated by state-of-the-art methodologies. \name{} introduces a communication interface that allows compilers and application developers to inform the hardware about true branch dependencies, confidential control-flow instructions, and fine-grained instruction constraints in order to apply restrictions only when necessary. We evaluate \name{} against known speculative execution attacks and in addition, present a new speculative fetch attack variant on the Pattern History Table (PHT) in branch predictors that shows  how similar previously reported vulnerabilities are more dangerous by enabling unprivileged attacks, especially with the state-of-the-art branch predictors. \name{} provides stronger security guarantees compared to the existing defenses while reducing the performance overhead of two state-of-the-art defenses from \DOLMAPerfOverhead{} and \sttPerfOverhead{} to just \SpecControlOverhead{}.

\end{abstract}

\maketitle

\pagestyle{plain}

\section{Introduction}

Speculative execution attacks, like Spectre~\cite{Spectre2019Kocher}, are a major concern in current and future modern processor designs as they exploit the main enabler of their performance, speculative execution~\cite{mcfarlin2013discerning}. Speculative execution attacks trick the processor into executing unintended paths of the program in a speculative manner and force the victim to access sensitive or secret information. The attacker can then reconstruct the secret data by probing the components containing the footprints of mispredicted and secret dependent paths which have not been rolled back after misspeculation. 
To mitigate speculative execution attacks, the most comprehensive solutions try to prevent transmitting secret information through any side-channel by restricting the execution of speculative instructions. 
For example, STT~\cite{yu2019speculative} deploys a dynamic taint tracking technique that restricts the execution of instructions that are tainted by speculative loads. Following solutions~\cite{loughlin2021dolma,choudhary2021speculative,schwarz2020context} use the same insight,
and propose hardware-only mechanisms to detect and restrict the speculative execution of problematic instructions.
These solutions provide secure speculation for \textit{sandboxed} programs that guarantee their memory accesses are within the authorized address range during the correct execution.
However, they fail to provide secure speculation for \textit{constant-time} programs, since these programs might load the secret data non-speculatively. To protect non-speculative secrets, speculative execution of all instructions tainted by secret values must be restricted. Current solutions for constant-time programs either manually specify secret regions of memory~\cite{fustos2019spectreguard,schwarz2020context,daniel2023prospect} or assume all memory regions are secret and declassify them only if they leak during non-speculative execution~\cite{choudhary2021speculative}. 

Unfortunately, most of these solutions tend to significantly reduce the benefits of speculative execution, as they follow a conservative approach of restricting the execution of the majority of speculative instructions after an unresolved branch.
Our study shows that STT and DOLMA incur \sttPerfOverhead{} and \DOLMAPerfOverhead{} performance overhead over an unprotected processor, respectively\footnote{We use modified versions of STT (to protect non-speculative secrets) and DOLMA (disabling Delay-on-Miss optimization to protect against speculative interference attacks~\cite{behnia2021speculative}). 
}. However, our studies show that
many of the restrictions introduced by these mitigations are unnecessary because not all instructions after a branch are truly dependent on the branch outcome. Such independent instructions can safely execute without compromising security. 
Hardware-only mitigations do not have sufficient information on true branch dependencies and all possible control-flow paths of the program, as the processor only views a small window of instructions at any given moment. 
Our key insight in this paper is that a hardware/software co-design provides a more efficient solution, where the software informs the hardware about true branch dependencies and the hardware applies restrictions only to truly speculative instructions.

In addition, prior solutions offer a limited scope of protection, as they only protect against current attacks exploiting speculative execution. However, modern processors have several components that hold speculative and confidential information (\textit{e.g.,} the Branch Prediction Unit) that can potentially leak, leading to the frequent discovery of new vulnerabilities.
Previous research has shown that the branch predictor can leak secret information through the Pattern History Table (PHT)~\cite{evtyushkin2018branchscope}. However, it was not considered a serious threat due to the limited leakage and attack assumptions. 
We demonstrate a new variant of the attack that an unprivileged attacker can extract a 19-bit private key of RSA without priming the PHT and we show that this attack becomes even more serious with better branch predictors.
We discuss that such attacks need to be categorized as speculation-based vulnerabilities because the secret extraction phase of the attack exploits the speculation operations at the processor's front-end. We call this class of attack \textit{Speculative Fetch Attacks}.
Speculatively fetching instructions and redirecting fetch in case of a misprediction creates the timing difference required to recover the control-flow information of the victim.

Defenses for speculative fetch attacks tend to be very challenging and inefficient, and often even impractical.
Software-only defenses require the program to remove all the confidential control-flow instructions (\textit{i.e.}, secret dependent branches or branches representing private activities of the users). 
Such transformation (\textit{e.g.}, if-converting the branches~\cite{mahlke1992effective}) is challenging for real-world applications, with respect to performance and practicality of removing complex control-flow patterns.
On the other hand, hardware-only solutions have no other choice other than disabling the branch predictor since they have no information about the confidentiality of newly fetched branches.
We believe that a hardware/software co-design and communication of confidential branches 
allows high-performance processors to enforce fine-grained front-end restrictions only for sensitive branches.

In this work, we propose \name{}, a mechanism to provide both comprehensive security and high performance through a configurable hardware/software co-design.
To accomplish this, we introduce a software interface for compilers and application developers to communicate true branch dependencies and confidential control-flow instructions to the hardware.
Our hardware uses software information to apply restrictions only if necessary, either at the front-end or the back-end of the processor
(mitigating both speculative fetch and execution attacks). 
\name{} provides secure speculation for both sandboxing and constant-time policies, in addition to
a new policy that guarantees the confidentiality of control-flow instructions (called 
\textit{control-flow confidentiality}).
The \name{} interface is flexible enough to 
define new secure speculation policies 
for future attacks and programs that might require additional or fewer restrictions, 
without additional changes to the hardware in the future.

The main contributions of this work are:
\begin{itemize}
    \item A novel hardware/software co-design methodology, called \name{}, that enables configurable and targeted mitigation for speculation-based attacks;
    \item Our targeted and automated mechanism reduces performance overhead compared to state-of-the-art solutions, 
    DOLMA~\cite{loughlin2021dolma} and STT~\cite{yu2019speculative}, from \DOLMAPerfOverhead{} and \sttPerfOverhead{} to just \SpecControlOverhead{},
    with negligible power and area overheads;
    \item We demonstrate a new variant of speculative fetch attacks that allows an unprivileged attacker to extract a 19-bit RSA private key from modern branch predictors. Additionally, we show the possibility of the attack on Intel, AMD, and Apple CPUs\footnote{We have responsibly disclosed our attack to the affected vendors and received the approval to distribute our findings.};
    \item Enhanced security against speculative fetch attacks.
\end{itemize}

\begin{figure}[t]
    \centering
    \includegraphics[trim=0.5cm 0.5cm 0 0.2cm,width=\linewidth]{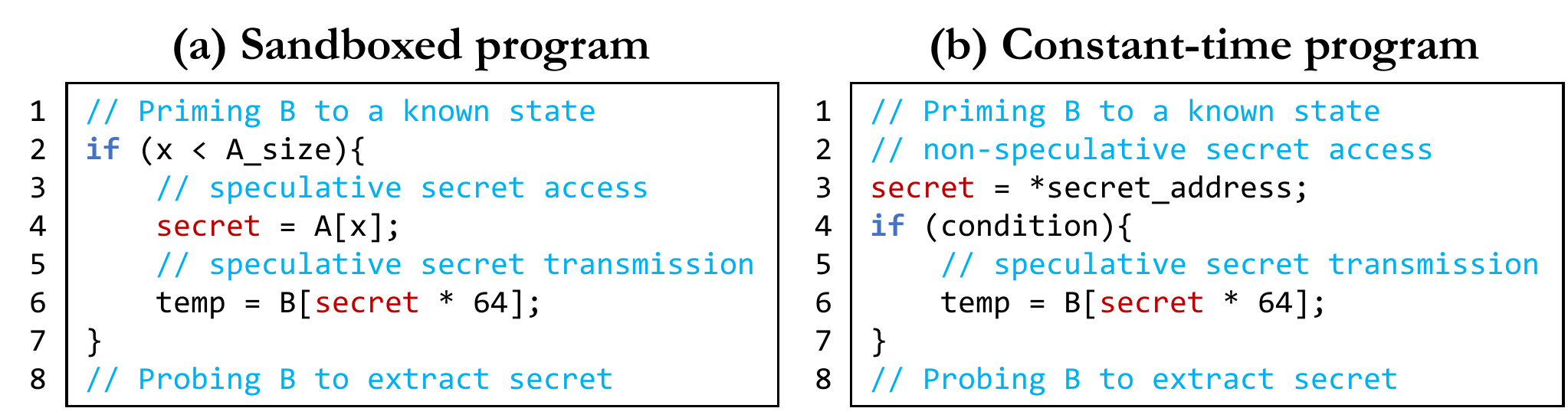}
    \caption{Victim gadgets for speculative execution attacks. 
    }
    \label{fig:spectre-gadgets}
\end{figure}

\section{Background and Motivation}
\label{sec:background}

In prior work~\cite{weisse2019nda,yu2019speculative,loughlin2021dolma,choudhary2021speculative,schwarz2020context}, an attack is considered speculative only if the secret is leaked at the processor back-end during speculative execution.
In this work, we expand this definition to consider an attack to be speculative 
if any step of the attack exploits speculative operations of the core to recover secret information,
even if the secret is transmitted to a channel non-speculatively. In this section, we describe two categories of speculation-based attacks:
(1) speculative execution, and (2) speculative fetch attacks.

\subsection{Speculative Execution Attacks}
\label{sec:spec-execution}

Speculative execution attacks %
aim
to bypass the address protection mechanisms of sandboxes or speculatively transmit the secret data of constant-time programs (see Figure~\ref{fig:spectre-gadgets}). %
In these attacks, the attacker (a) mistrains the branch predictor to always take the victim branch even if the condition is not true (line 2 in Figure~\ref{fig:spectre-gadgets}a and line 4 in Figure~\ref{fig:spectre-gadgets}b), and (b) speculatively (loading the secret into the core in a sandboxed program and) transmitting the secret to a primed side-channel (\textit{e.g.,} the cache). The processor then detects the branch misprediction, rolls back the misspeculated state, and resumes the execution from the correct path. However, the transmitted secret leaves behind persistent state that the attacker can use to extract the secret. Spectre-V1~\cite{Spectre2019Kocher} was the first attack exploiting this vulnerability, with several follow-up variants~\cite{NetSpectre2018Schwarz,SmotherSpectrePortContention2019,chen2019sgxpectre,kiriansky2018speculative,koruyeh2018spectre,maisuradze2018ret2spec,horn2018speculative}.

\subsubsection{Existing Defenses}

Many initial defenses for speculative execution attacks~\cite{yan2018invisispec,Bourgeat2019MI6,khasawneh2019safespec,saileshwar2019cleanupspec,qureshi2018ceaser,DOM,Omar2021ironhide} focused on solutions to mitigate specific channels, such as caches~\cite{Spectre2019Kocher,SpectreReturns2018Koruyeh,SpectreVariantCFH2019Mambretti,Kiriansky2018SBOSpectre,kirzner2021analysis}.
However, new Spectre variants have demonstrated that other components in the core can act as a side-channel as well, such as the Branch Target Buffer (BTB)~\cite{weisse2019nda}. As an attempt to build a more comprehensive set of countermeasures, recent works have focused on mitigating the attack at the source~\cite{yu2019speculative,SDOSTT2020Yu,yu2019data,weisse2019nda,schwarz2020context,loughlin2021dolma,daniel2023prospect}, by completely restricting speculative execution of instructions that can potentially reveal sensitive information. The focus of these works is to block leaks through any potential channel, but only for speculative execution attacks. 
However, there are still challenges to be addressed as hardware-only solutions fail to completely solve. 

\subsection{Speculative Fetch Attacks}
\label{sec:spec-fetch}

\begin{figure}[t]
    \centering
    \includegraphics[trim=0 .95cm 0 1cm,width=1\linewidth]{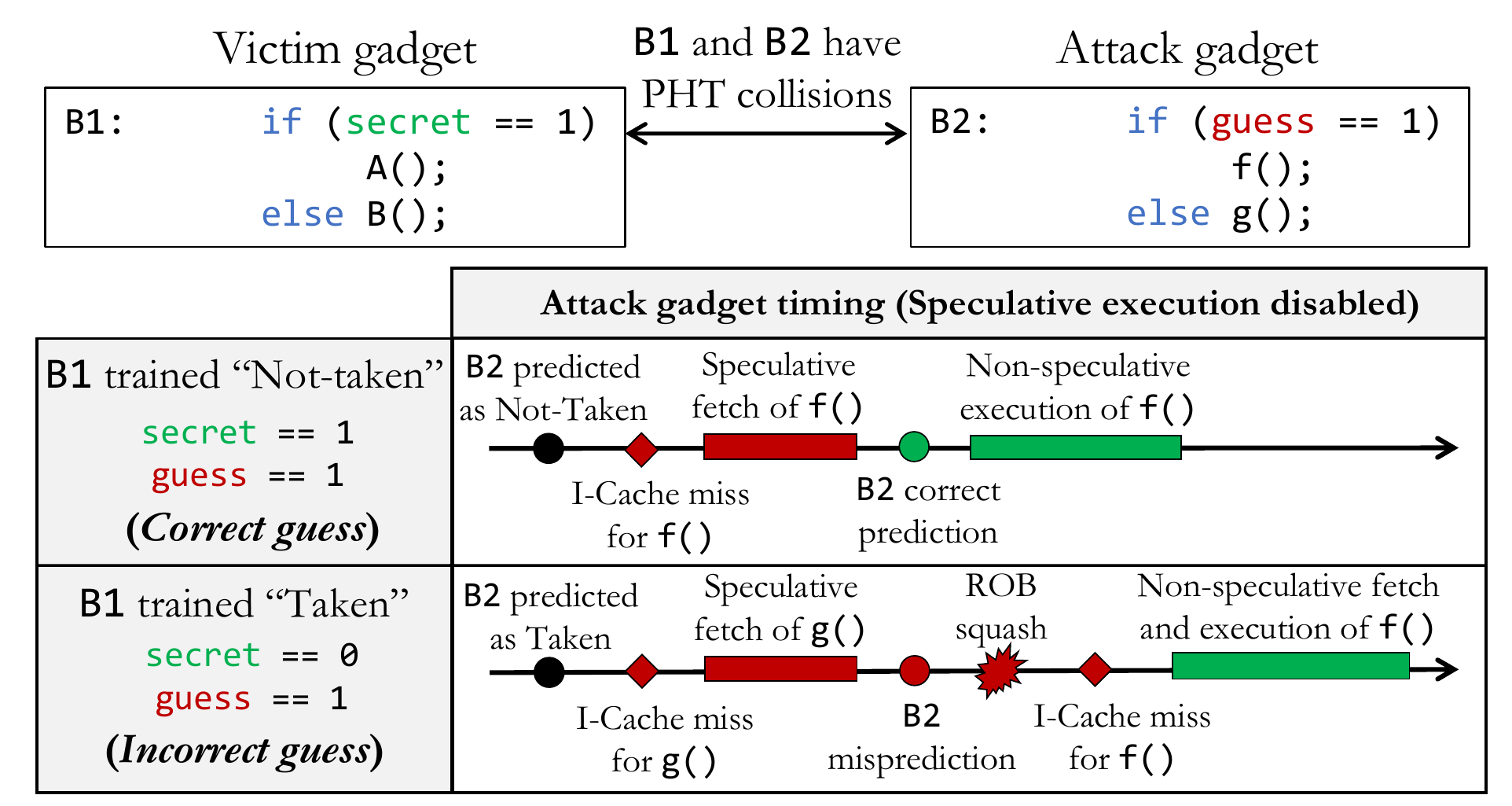}
    \caption{Events and timing of correct prediction (correct guess) and misprediction (incorrect guess) in a speculative fetch attack.
    We assume that before executing the attack gadget all instructions and data of the attacker are flushed from the caches. Note that, Not-Taken means executing the fall-through path (\texttt{A()} and \texttt{f()} in this example).}
    \label{fig:gadget-events}
\end{figure}

Speculative fetch attacks exploit the fact that the branch predictor remembers the decisions originally made by the victim. 
Figure~\ref{fig:gadget-events} shows a victim gadget that consists of a confidential branch that is vulnerable to speculative fetch attacks (Branch \texttt{B1}).
BranchScope~\cite{evtyushkin2018branchscope} is an example of such attack, where the attacker performs three steps to extract a secret key: (1) the attacker primes one of the PHT entries into a known state (\textit{e.g.,} strongly taken), (2) the victim executes its secret dependent branch (colliding with the primed PHT entry), and (3) the attacker can extract the secret bit by probing the same PHT entry. Unfortunately, BranchScope requires OS control to interrupt the victim exactly at each iteration of the secret dependent branch to extract the secret bit by bit.
We consider this attack to be a speculation-based vulnerability as the secret extraction happens through speculative predictions initiated by the branch predictor at the fetch stage.

\subsubsection{A New Variant of Speculative Fetch Attacks}
\label{sec:conjuring}

We present a new variant of speculative fetch attacks that, unlike previous work~\cite{evtyushkin2018branchscope}, allows an \textit{unprivileged} attacker to successfully launch an attack, \textit{without requiring the OS to precisely control the victim execution} and \textit{without requiring the attacker to prime the branch predictor}. 
We exploit the fact that state-of-the-art branch predictors can remember branch patterns 
and therefore can extract the confidential information that has already been learned.
Our attack only performs two steps without interfering with the victim's execution:

    \textbf{Step 1 (victim): Train the PHT with the secret}. The victim runs normally and trains the PHT with its secret information and branch decisions;
    
    \textbf{Step 2 (attacker): Measure the execution time of the attack gadget}. After victim execution, the attacker needs to (a) create a PHT collision with the victim branch~\cite{canella2019systematic,kirzner2021analysis} and (b) make a guess about the secret. 
    The attacker evaluates the correctness of the guess to extract the secret. 

\begin{figure}[t]
    \centering
    \includegraphics[trim=0 0.8cm 0 0.5cm,width=0.85\linewidth]{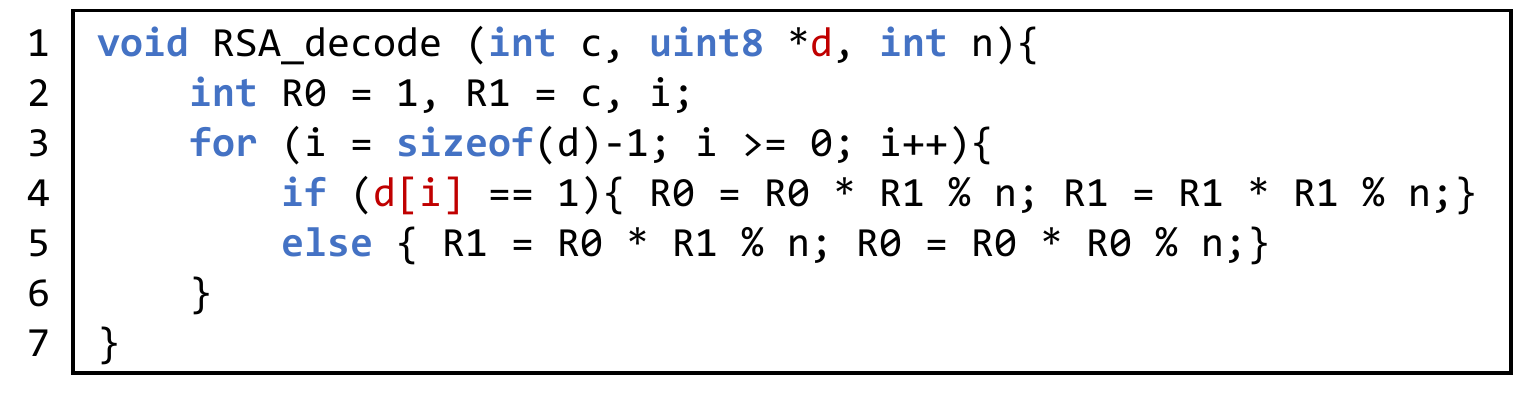}
    \caption{RSA modular exponentiation using Montgomery Ladder Powering~\cite{joye2002montgomery,opensslrsaml}.}
    \label{fig:RSA-decode}
\end{figure}

\begin{figure}[t]
    \centering
    \includegraphics[trim=0 0.9cm 0 0.5cm,width=\linewidth]{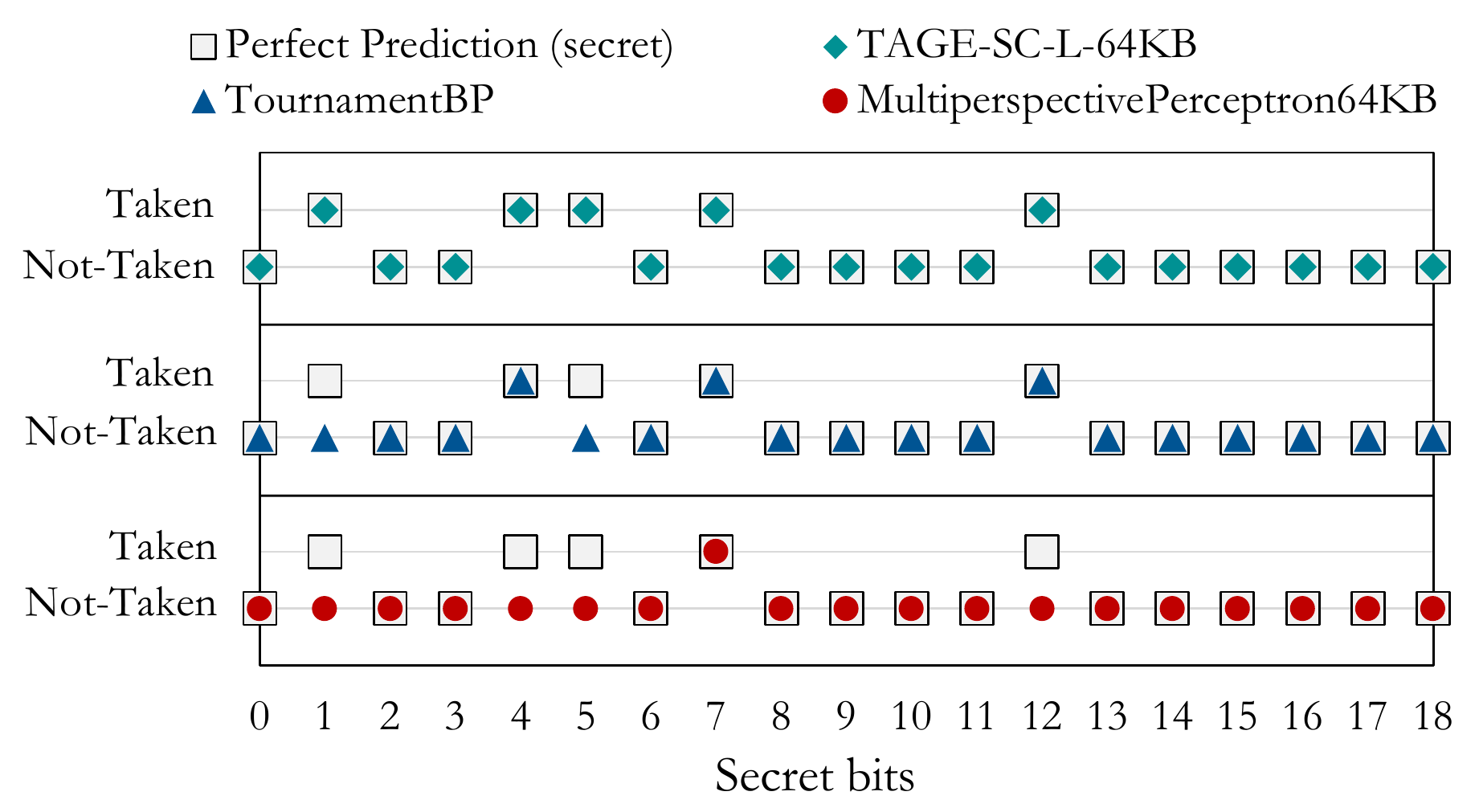}
    \caption{Investigation of different branch predictors in gem5 and their prediction for RSA private key after 1,000 rounds of decryption. The results show that TAGE-SC-L-64KB remembers the entire key. 
    Note, that a Taken decision means that the secret bit is 0 based on the RSA code in Figure~\ref{fig:RSA-decode}. The secret is $(1011001011110111111)_{2}$.}
    \label{fig:bp_main}
\end{figure}

Figure~\ref{fig:gadget-events} shows an example of victim and attack gadgets, along with a timeline of the events during the second step of the attack.
If the attacker guesses correctly, the instructions are fetched from the correct path and are not squashed later. However, in the case of an incorrect guess, the incorrectly fetched instructions are squashed and the processor starts fetching and executing instructions from the correct path (speculative execution is disabled in Figure~\ref{fig:gadget-events}). As shown in the figure, the attacker can evaluate the correctness of the guess by measuring the execution time of the attack gadget.

This attack exploits the victim itself to prime and train the PHT state with the secret. Our investigations using gem5 simulation~\cite{binkert2011gem5} with different branch predictors (Figure~\ref{fig:bp_main}) show that TAGE-SC-L-64KB~\cite{CBP}, the current state-of-the-art branch predictor, is able to learn all 19 bits of the private exponent in a Montgomery Ladder RSA~\cite{joye2002montgomery} (Figure~\ref{fig:RSA-decode}) after 1,000 rounds of decryption. 
Older branch predictors fail to remember all the bits. For example, MultiperspectivePerceptron64KB~\cite{jimenez2001dynamic,jimenez2003fast} provides incorrect predictions for 4 of the secret bits.
In other words, deploying increasingly better branch predictors make speculative fetch attacks more serious than was previously realized.
In Section~\ref{sec:real-attacks}, we demonstrate the feasibility of our attack on real existing CPUs.

\subsubsection{Attack Surface}

Any program with control-flow decisions that carry confidential information is vulnerable to speculative fetch attacks.
Many existing cryptographic implementations have secret dependent branches, like the RSA modular exponentiation in GnuPG 1.4~\cite{gnupg}, OpenSSL v1.01e~\cite{opensslrsaml}, MbedTLS 3.0.0~\cite{mbed}, LibreSSL 3.3.6~\cite{libressl}, and wNAF algorithm in secp256k1 curve (found in OpenSSL v1.1.0h~\cite{fan2016attacking}). 
In addition, many real-world applications have branches that can leak confidential information (\textit{e.g.}, users' Bluetooth connections~\cite{bluetooth} and battery properties~\cite{battery} in the Linux kernel). 

\subsubsection{Existing Defenses}
\label{sec:spec-fetch-defenses}

To defend against speculative fetch attacks, software-only defenses %
adopt
strict policies like data-oblivious programming~\cite{schwarzl2021specfuscator,liu2015ghostrider,molnar2006program,rane2015raccoon} (\textit{e.g.}, removing secret dependent branches).
Adapting to such guidelines is challenging for real-world and control-flow intensive programs, and even many cryptographic modules still choose to use secret dependent branches. These applications secure non-speculative leaks of the program with lightweight spot defenses based on the adversaries' capabilities\footnote{For example, a minor compiler update mitigates the fetch bandwidth misalignments exploited by Frontal attacks~\cite{puddu2021frontal}.}. 
On the other hand, hardware-only defenses have no prior information about newly fetched branches at the front-end and their only option is to disable the branch predictor, an extreme measure that significantly reduces %
performance, especially in highly speculative modern processors.
We show that, for example, a secure baseline processor with no branch predictor can introduce
up to a $3.72 \times$ performance overhead.
Note, that clearing the branch predictor after context switches does not solve the problem for hyperthreading-based scenarios and also requires the trust of the OS to not exploit the branch predictor's state before clearing it.

This speculative fetch vulnerability is a microarchitectural leak that software verification tools cannot reason about (as not all of the microarchitectural features of processors are disclosed); they can only analyze the correct execution of the program.
In this work, we introduce a new secure speculation policy, called \textit{control-flow confidentiality}, %
where
the hardware guarantees secure speculation for confidential branches.

\subsection{Challenges of the State-of-the-Art}

In this work, we identify four challenges of state-of-the-art defenses for speculation-based vulnerabilities:

\textbf{(1) \textit{Hardware-only solutions for speculative execution attacks apply unnecessary restrictions for the instructions that will not leak data due to speculation}}. Our analysis shows that not all instructions after unresolved branches are truly dependent on the branch~\cite{hajiabadi2021noreba}. 

\textbf{(2) \textit{Software-only and hardware-only solutions cannot efficiently protect confidential control-flow decisions}}.
Existing mechanisms are challenging for real-world applications and incur prohibitive performance overheads.

\textbf{(3) \textit{Future security challenges}}:
The rapid development of new speculative attacks~\cite{barberis2022branch,wikner2022retbleed,behnia2021speculative,li2022unxpec} invalidates many of the existing state-of-the-art (performance-wise and security-wise) defenses, and they are unable to recover from newly discovered attacks.
Hence, application developers and CPU vendors currently need to fall back to heavy-handed protections (\textit{e.g.}, fences, or strict programming guidelines).
A fine-grained solution providing sufficient protections that limits performance penalties would help to alleviate these issues. 

\textbf{(4) \textit{Future performance challenges}}:
While new speculation vulnerabilities emerge rapidly, our understanding of the programs and vulnerable workloads can continue to improve. 
We believe that even a hardware/software co-design, as we propose in this work, needs to be flexible enough to benefit from future compiler and programming enhancements.
For example, future compilers might be able to extract more accurate and less conservative instruction dependencies or new efficient programming disciplines that require fewer restrictions from the hardware. 

\subsection{Our Approach: HW/SW Co-design}
Our key insight is that we can enable fine-grained and targeted restriction of speculative fetch and speculative execution only for problematic instructions 
using targeted %
hardware/software communication.
In this work, we propose a software interface that allows compilers and application developers to communicate confidential branches (to restrict them at the front-end and defend against speculative fetch attacks) and true branch dependencies (to only restrict the execution of the instructions that are truly dependent on unresolved branches and defend against speculative execution attacks).
Our proposed microarchitecture only modifies the front-end logic to detect and restrict the fetch of confidential branches and adds a new table at the backend, the Unresolved Branches Table (UBT), that keeps track of the live unresolved branches residing in the ROB and applies execution restrictions to an instruction only if it depends (as informed by the compiler) on any of the branches in the UBT.

Our hardware/software co-design is flexible enough to allow compilers and developers to apply or relax front-end and back-end restrictions without
additional hardware updates
to provide
(1) a \textit{configurable} methodology to defend against future vulnerable instructions and paths of the programs, and (2) the flexibility to configure the hardware restrictions based on future advancements of programs and compilers, hence, providing a potential for future performance improvements. 

\begin{figure*}[t]
    \centering
    \includegraphics[trim=0 0.5cm 0 1cm,width=0.92\linewidth]{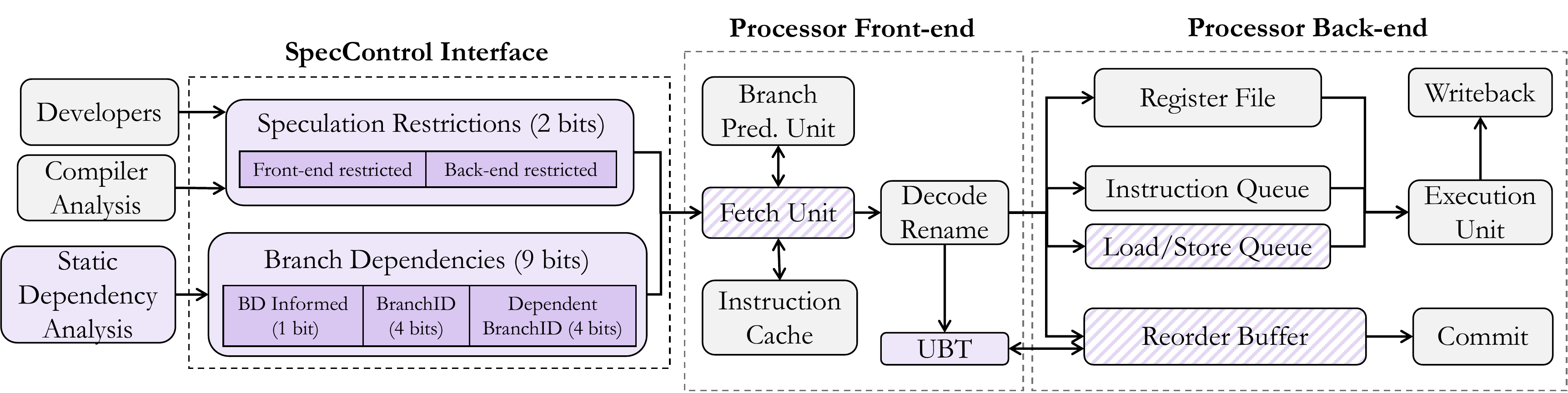}
    \caption{\name{} design. The highlighted components are added by \name{} on top of the baseline compiler and processor and the hashed components are modified by \name{}. UBT is the Unresolved Branches Table that keeps the live unresolved branches in the ROB.
    Note, that the Speculation Restrictions and Branch Dependencies are set per instruction through prefix bytes of x86 ISA (11 bits required for each instruction).
    The Load/Store Queue is hashed, similarly to prior work~\cite{weisse2019nda,yu2019speculative,loughlin2021dolma,daniel2023prospect}, we updated it to prevent store-to-load bypass attacks~\cite{horn2018speculative}.}
    \label{fig:design}
\end{figure*}

\section{Threat Model}
\label{sec:threat_model}

In this paper, 
we focus on securing speculation for software following (1) sandboxing policies, (2) constant-time policies~\cite{guarnieri2021hardware,daniel2023prospect}, as well as (3)  control-flow confidentiality, covering all known
\textit{speculative fetch}, and \textit{speculative execution} attacks (see Section~\ref{sec:sec-analysis} for the definitions and analysis of our security guarantees).  

Speculative execution attacks include all Spectre-like attacks~\cite{Spectre2019Kocher,NetSpectre2018Schwarz,SmotherSpectrePortContention2019,chen2019sgxpectre,kiriansky2018speculative,koruyeh2018spectre,maisuradze2018ret2spec,horn2018speculative}.
Speculative fetch attacks refer to speculation vulnerabilities introduced at the front-end, like BranchScope~\cite{evtyushkin2018branchscope} and our new variant of the attack.
The assumptions we make for speculative fetch attacks are: 
(1) the Pattern History Table (PHT) must be shared between the attacker and the victim. This means that the attacker and the victim must run on the same physical core and share PHT entries,
(2) the attacker can trigger or is able to watch for and validate the victim code execution, and (3) 
the attacker can launch the attack with user-only privileges in the system, without requiring any OS capabilities. 

\textbf{Out of Scope:} We do not consider Meltdown-type attacks~\cite{Lipp2018meltdown,van2018foreshadow,weisse2018foreshadow,van2019ridl,canella2019fallout,schwarz2019zombieload} 
that exploit the delay between exceptions and when they are handled. 
New processors patch this class of attacks~\cite{mds}.
In addition, attacks exploiting non-speculative leaks of the programs~\cite{puddu2021frontal,lee2017inferring,lipp2022amd,gruss2016prefetch,vicarte2022augury,chen2023afterimage} are out of scope of this work. 

\section{\name{} Design}
\label{sec:design}

\name{} design deploys a configurable solution that efficiently applies minimal restrictions to guarantee strong protections.
It provides a lightweight software interface (Section~\ref{sec:software}) that allows the compiler and the applications developers to mark true branch dependencies for targeted front-end or back-end restrictions. 
This information is communicated to the hardware and the processor applies the proper restrictions for marked instructions (Section~\ref{sec:uarch}). 
Figure~\ref{fig:design} shows the \name{} design, highlighting the modifications over existing processor designs and compilers.
\name{} has always-on security for sandboxed and constant-time programs, even for legacy binaries not compiled by our compiler. But, if needed, \name{} protections can be disabled by simply instrumenting the binary and manipulating the restriction bits of the x86 instruction prefix bits.
We built and verified the end-to-end and automated design of \name{} in LLVM compiler~\cite{lattner2004llvm} and gem5 simulation~\cite{binkert2011gem5}.

\subsection{\name{} Interface}
\label{sec:software}

\name{} interface has two main features: (1) \textit{marking true branch dependencies of the instructions} (Section~\ref{sec:dependencies}); this will help the processor to only restrict speculative execution of the instructions that are truly dependent on unresolved branches, (2) \textit{Marking the confidential control-flow instructions} (Section~\ref{sec:restrictions}); this informs the processor to restrict the fetch unit upon marked instructions. While these two features mitigate both speculation execution and speculative fetch attacks, \name{} interface is highly configurable to apply or relax other fine-grained restrictions. 

\subsubsection{Communicating Branch Dependencies}
\label{sec:dependencies}

While the hardware assumes all instructions after an unresolved branch are control dependent, a static compiler analysis shows that not all instructions are truly dependent on the branch's outcome.
Therefore, restricting the execution of all instructions after unresolved branches is not required to provide secure speculation; since independent instructions do not leave unintended persistent changes if a branch is mispredicted. Current hardware designs, however, cannot reason about true branch dependencies, as they do not have information about the entire program and its control flow~\cite{hajiabadi2021noreba}. 

\begin{table*}[t]
  \centering
  \scalebox{0.75}{
  \begin{tabular}{c|c|c|c||c|c|c|c||c|c|c|c}
  \toprule
  \cellcolor{gray!10}
  \textbf{Tag} & \rotatebox[origin=c]{90}{\parbox{1.5cm}{\centering \textbf{Front-end}}} & \rotatebox[origin=c]{90}{\parbox{1.5cm}{\centering \textbf{Back-end}}} & \textbf{Action} & 
  \cellcolor{gray!10}
  \textbf{Tag} & \rotatebox[origin=c]{90}{\parbox{1.5cm}{\centering \textbf{BD Informed}}} & \rotatebox[origin=c]{90}{\parbox{1.5cm}{\centering \textbf{Dependent\\BranchID}}}  & \textbf{Action} & 
  \cellcolor{gray!10}
  \textbf{Tag} &  \rotatebox[origin=c]{90}{\parbox{1.5cm}{\centering \textbf{BD Informed}}} & \rotatebox[origin=c]{90}{\parbox{1.5cm}{\centering \textbf{BranchID}}} & \textbf{Action} \\
  
  \midrule
  
  \cellcolor{gray!10}
   & \multirow{2}{*}{0} & \multirow{2}{*}{0} & \multirow{2}{*}{No additional restrictions} & 
   \cellcolor{gray!10}
    & \multirow{2}{*}{X} & \multirow{2}{*}{INVALID} & The instruction will be dependent & 
   \cellcolor{gray!10}
    & \multirow{2}{*}{X} & \multirow{2}{*}{INVALID} & The branch will restrict all \\
   
   \multirow{-2}{*}{\cellcolor{gray!10} \resNo{}} &  &  &  & 
   \multirow{-2}{*}{\cellcolor{gray!10} \bdInvalid{}}&  &  & on the most recent branch$*$ &
   \multirow{-2}{*}{\cellcolor{gray!10} \brInvalid{}}& &  & instructions afterwards$*$ \\
   
   \hline

   \cellcolor{gray!10}
    & \multirow{2}{*}{1} & \multirow{2}{*}{0} & \multirow{2}{*}{Restricted at the front-end$\ddagger$} & 
   \cellcolor{gray!10}
    & \multirow{2}{*}{0} & \multirow{2}{*}{VALID} & \multirow{2}{*}{No branch dependency restriction} & 
   \cellcolor{gray!10}
    & \multirow{2}{*}{0} & \multirow{2}{*}{VALID} & The branch will not restrict \\
   
   \multirow{-2}{*}{\cellcolor{gray!10} \resFE{}} &  &  &  & 
   \multirow{-2}{*}{\cellcolor{gray!10} \bdNo{}} &  &  &  & 
   \multirow{-2}{*}{\cellcolor{gray!10} \brNo{}} & &  & any instruction afterwards \\
   
   \hline
   
   \cellcolor{gray!10}
    & \multirow{2}{*}{0} & \multirow{2}{*}{1} & \multirow{2}{*}{Restricted at the back-end} & 
   \cellcolor{gray!10}
    & \multirow{2}{*}{1} & \multirow{2}{*}{VALID} & The instruction will be dependent & 
   \cellcolor{gray!10}
    & \multirow{2}{*}{1} & \multirow{2}{*}{VALID} & The branch will restrict \\
   
   \multirow{-2}{*}{\cellcolor{gray!10} \resBE{}} & & & & 
   \multirow{-2}{*}{\cellcolor{gray!10} \bdValid{}} & & & the branch specified by BranchID &
   \multirow{-2}{*}{\cellcolor{gray!10} \brValid{}} & & & the dependent instructions \\
  \bottomrule
  \end{tabular}
  }
  \caption{\name{} options to mark restrictions for an instruction. $\ddagger$ Note, that setting both front-end and back-end bits means restricting at both stages, however, it has the same functionality as \resFE{}; because if an instruction is restricted at the front-end then the next instructions will be non-speculative at the back-end and no extra restriction will be applied. $*$ This policy ensures secure speculation for legacy binaries (except for control-flow confidentiality).}
  \label{table:restrictions}
\end{table*}

Algorithm~\ref{algo:traverse} shows how we detect all the dependent instructions of a conditional branch that performs a traversal search over the data-flow graph (DFG).
First, we initialize the \texttt{working\_set} with the branch control dependent instructions (\textit{i.e.}, all the instructions reachable between the branch and its reconvergence point; line 3). Next, we visit all the direct dependent instructions of the \texttt{working\_set} (line 9). Direct dependencies are determined through the compiler def-use and alias analysis. 
If two instructions have a may-alias dependency we consider it as a dependency in order to provide a fully non-speculative and conservative list of dependencies.
At the end of this search, all the dependent instructions will be in the \texttt{dependents} set of the input branch (line 10). 
This compiler pass statically considers all possible paths of the program after a branch and conservatively considers all may-alias dependencies as true dependencies, hence, it is sound by design and does not miss any potential dependency.

\begin{algorithm}[t]
\footnotesize
\DontPrintSemicolon
\SetAlgoLined
\SetKw{KwInit}{Initialization}
\KwIn{Branch $BR$}
\KwOut{Determining all the dependents of $BR$}
$working\_set$.clear()\\
$processed\_insts$.clear()\\
$working\_set \gets BR.control\_dependents$\\
\While{$\neg working\_set$.empty()}{
    $inst$ $\gets$ $working\_set$.pop()\\
    \If{$inst \in processed\_insts$}{
        continue\\
    }
    \ForEach{$dep \in inst$.direct\_dependents}{
        $BR.dependents$.insert($dep$)\\
        $working\_set$.push($dep$)\\
    }
    $processed\_insts$.insert($inst$)\\
}
\caption{Branch Dependents Traversal}
\label{algo:traverse}
\end{algorithm}

\textbf{Marking dependencies.} First, we assign a static identifier (BranchID) to each branch
(see \texttt{BranchID} in Figure~\ref{fig:design}). 
Second, we mark each instruction with the BranchID of the most recent dependent branch detected in the previous step (see \texttt{Dependent BranchID} in Figure~\ref{fig:design})\footnote{In the rare case of an instruction depending on multiple independent branches, we either need to communicate multiple branches or conservatively assume that the instruction depends on all prior unresolved branches 
(similar to state-of-the-art hardware-only solutions~\cite{loughlin2021dolma}), by marking the branch as \brInvalid{}.
}. We embed this information in the prefix bytes of the x86 ISA instructions. 
Note, that other ISAs like RISC-V can define new instructions to communicate the compiler information but at the cost of extra instruction decoding.
Our experiments show that 4 bits are enough to represent the BranchID. In addition, we set the \texttt{BD Informed} bit to 1 for the branches that our compiler is providing dependency information. This bit is used to (1) support both legacy binaries and (2) allow disabling protections if desired (see Table~\ref{table:restrictions} for detailed options).

\subsubsection{Communicating Speculation Restrictions}
\label{sec:restrictions}

The \name{} interface also allows the developers and the compiler to directly mark instructions that need to be
restricted either at the front-end or the back-end of the processor.
In our baseline x86 ISA, we use two prefix bits of the instructions to inform the hardware about these marked restrictions:
one bit that indicates if the instruction needs to be restricted at the front-end (\texttt{Front-end restricted} in Figure~\ref{fig:design}) and one bit to restrict the instruction at the back-end (\texttt{Back-end restricted}).
Table~\ref{table:restrictions} shows all restriction markings in \name{}. 
\resFE{} refers to front-end restriction and \resBE{} means restricted at the back-end. \resNo{} also indicates no restriction.

To address speculative fetch attacks, we need to mark secret dependent and confidential branches as \resFE{}, which restricts the branch at the front-end from accessing/updating the branch predictor.
For the evaluation of this work, we deploy a state-of-the-art taint tracking implementation~\cite{borrello2021constantine} to mark secret dependent branches in a set of realistic cryptographic modules~\cite{borrello2021constantine,rane2015raccoon,liu2015ghostrider,wu2018eliminating}.
However, there are several ways to mark secret dependent branches depending on the application. Either the developer can directly mark them by using OpenMP-like programming directives~\cite{chandra2001parallel}, or the compiler can automatically detect secret dependent branches using compiler analysis, as we used in this work. 

\begin{table*}[t]
  \centering
  \scalebox{0.87}{
  \begin{tabular}{cc|l}
  \toprule
  
  & \textbf{Event} & \multicolumn{1}{c}{\textbf{Action}}\\
  
  \midrule

  \multirow{2}{*}{\step{1}} & Branch $BR$ & while $BR$ is unresolved:\\
  & (marked as \resFE{}) fetched & \hspace{0.1 in} block fetch\\

  \hline

  \multirow{2}{*}{\step{2}} & Instruction $Inst$ & \multirow{2}{*}{$ROB[Inst].BackendRestricted = 1$} \\
  & (marked as \resBE{}) enters ROB & \\

  \hline

  \multirow{2}{*}{\step{3}} & Instruction $Inst$ & \multirow{2}{*}{$ROB[Inst].BackendRestricted = 0$} \\
  & (marked as \resBE{}) resolves & \\

  \hline
  
  \multirow{3}{*}{\step{4}} & Branch $BR$ & while UBT is full:\\
  & (marked as \brValid{}) & \hspace{0.1 in} stall inserting new instructions to the ROB\\
  & enters ROB & $UBT[BR.BranchID]=SeqNum$\\

  \hline
  
  \multirow{5}{*}{\step{5}} & & if $UBT[Inst.DependentBranchID]$ exists:\\
   & Instruction $Inst$ & \hspace{0.1 in} $ROB[Inst].Restricted = 1$\\
  & (marked as \bdValid{}) & \hspace{0.1 in} $ROB[Inst].DependentBranch=<Inst.DependentBranchID, UBT[Inst.DependentBranchID]>$\\
  & enters ROB & if an unresolved branch marked as \brInvalid{} or indirect jump not marked as \brNo{} exists in ROB:\\
  & & \hspace{0.1 in} $ROB[Inst].BackendRestricted = 1$\\

  \hline
  
  \multirow{4}{*}{\step{6}}& & for all entries $E$ in ROB:\\
    & Branch $BR$ & \hspace{0.1 in} if $ROB[E].DependentBranch == <BR.BranchID, BR.SeqNum>$:\\
  & (marked as \brValid{}) resolves& \hspace{0.2 in} $ROB[E].Restricted = 0$\\
  & & Remove $BR$ from UBT\\
  
  \hline
  
   \multirow{5}{*}{\step{7}} & & for all entries $E$ in ROB after $Inst$:\\
    & Branch $Inst$ (marked as \brInvalid{})& \hspace{0.1 in} if ($E$ is marked as \brInvalid{}) or ($E$ is indirect jump and marked as \brValid{}):\\
  & or indirect jump $Inst$  & \hspace{0.2 in} $break$\\
  & (marked as \brValid{} or \brInvalid{}) resolves & \hspace{0.1 in} else if $E$ not marked as \resBE{}:\\
   & & \hspace{0.2 in} $ROB[E].BackendRestricted = 0$\\
  
  \bottomrule
  \end{tabular}
  }
  \caption{Events and actions in \name{}. For instruction $Inst$, $Inst.DependentBranchID$ is specified by the compiler in the prefix bits and for branch $BR$, $BR.BranchID$ is an ID assigned by the compiler and specified in the prefix bits as well (see Figure~\ref{fig:design}).}
  \label{table:events}
\end{table*}

\subsection{\name{} Microarchitecture}
\label{sec:uarch}

The requirements for hardware implementation of \name{} are: (1) to collect the software information communicated by \name{} interface,
(2) to prevent the speculative fetch of the instructions that are marked as confidential (\textit{i.e.}, \resFE{}), and (3) to prevent speculative execution of the instructions that are dependent on at least one unresolved branch in the ROB or directly marked as back-end restricted (\textit{i.e.}, \resBE{}).  
Table~\ref{table:events} shows a detailed event-action description of the \name{} microarchitecture.

\subsubsection{Front-end Restrictions}

The fetch unit is modified to check and validate the speculation restriction bits of a fetched instruction. If the restriction bits of a branch are marked as \resFE{}, the fetch unit does not use the branch predictor to speculatively fetch subsequent instructions and instead stalls. When the branch resolves, the processor resumes fetching instructions from the correct, non-speculative path (step~\step{1} in Table~\ref{table:events}). 
This way, unsafe branches will not leave traces in the speculative components of the front-end (\textit{e.g.}, the branch predictor); hence, protecting against speculative fetch attacks.

\subsubsection{Back-end Restrictions}

\begin{figure}[t]
    \centering
    \includegraphics[trim=0 1.5cm 0 1cm,width=0.92\linewidth]{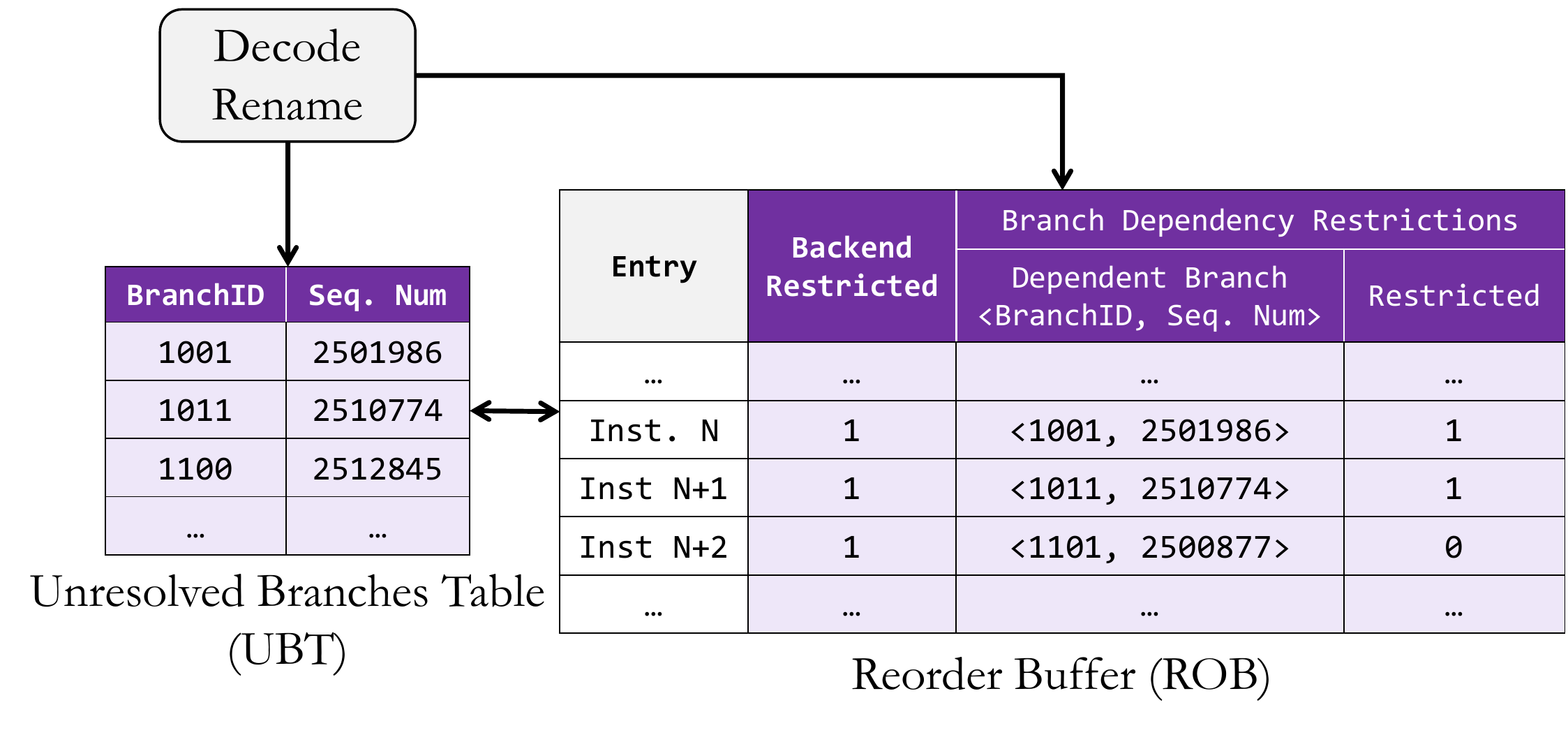}
    \caption{New structures used by \name{}. The Unresolved Branches Table (UBT) stores the speculative (unresolved) branches and maps the compiler-specified BranchID to the dynamic sequence number of the branch. ROB is modified to specify the most recent dependent branch of the instruction 
    and the restriction based on that branch, as well as one bit to indicate if the execution is directly restricted by the \resBE{} marking.}
    \label{fig:tables}
\end{figure}

For back-end restrictions, we add an extra bit to the ROB, called \texttt{Backend Restricted} (see Figure~\ref{fig:tables}).
If an instruction marked by the compiler as \resBE{} enters the ROB then the execution of the instruction will be restricted 
(\texttt{Backend Restricted} bit set to 1 in the ROB, step~\step{2}).
When the instruction becomes safe (\textit{e.g.}, all input operands are non-speculative) and guaranteed to commit (\textit{e.g.}, for loads and stores it means they resolve in Load/Store Unit and their page-table access succeeds) then the \texttt{Backend Restricted} bit is set to 0 and the instruction is allowed to execute and forward its results to the dependent instructions (step~\step{3}).

In addition, the \texttt{Backend Restricted} bit is used to restrict the instructions that are not analyzed by our compiler (\textit{e.g.}, instructions marked as \bdInvalid{} and branches marked as \brInvalid{}).
All instructions after a branch marked as \brInvalid{} will be restricted at the back-end until the branch resolves (step~\step{7}).
Hence, we provide always-on security for speculative execution attacks. The developers and the compiler can disable the protections by marking the instructions as \bdNo{} and \brNo{}, as explained in Table~\ref{table:restrictions}\footnote{Legacy binaries can also perform this by instrumentation.}.

\subsubsection{Branch Dependency Restrictions}

To correctly apply branch dependency restrictions communicated by the compiler, we introduce a new hardware structure, called the Unresolved Branches Table (UBT). The UBT stores all live unresolved branches (see Figure~\ref{fig:tables}). The ROB is also modified to: (1) indicate if the execution of an instruction is restricted due to branch dependency (\texttt{Restricted} bit in Figure~\ref{fig:tables}) and (2) store the Branch ID of the most recent dependent branch (\texttt{Dependent Branch} in Figure~\ref{fig:tables}).

When a branch instruction is decoded it will enter the ROB and update the UBT with its compiler-specified BranchID and unique dynamic sequence number (dynamic identifier of instructions) (step~\step{4}).
Note, that the branch should have \brValid{} marking which means it has been analyzed by the compiler.
In case of a full UBT, we stall inserting new instructions to ROB (\textit{i.e.}, ROB full) until at least one of the prior branches resolves and then resume decoding and inserting instructions to the ROB (see Section~\ref{sec:ubt-study} for the impacts of UBT size).
For every decoded instruction entering the ROB, the processor checks if the instruction's most recent dependent branch exists in the UBT. In case of a hit in the UBT, the instruction's \texttt{Restricted} bit is set to 1 in the ROB and the execution of the instruction will be restricted. The ROB is also updated with the BranchID and sequence number of the branch that the instruction depends on (step~\step{5}). 
Finally, when a branch resolves and the correct path is determined, the \texttt{Restricted} bit of all dependent instructions is set to 0 and they can execute (\textit{i.e.,} they are guaranteed to commit and not squash due to branch misprediction) (step~\step{6}). Note, that an instruction is allowed to execute only if both \texttt{Restricted} and \texttt{Backend Restricted} bits are 0.

For indirect jumps (\textit{e.g.,} function calls), all instructions after the jump need to be restricted at the back-end (step~\step{5}) and execute only if the jump resolves and becomes non-speculative (step~\step{7}), to address attacks like Spectre-BTB (V2) and it successors.
In \name{}, we restrict all instructions after an indirect jump unless it is marked as \brNo{} (protection disabled by the applications).

Similarly to prior work~\cite{loughlin2021dolma,yu2019speculative}, 
we prevent speculative fetch redirections to mitigate the attacks that exploit port contention in simultaneously multi-threaded (SMT) processors~\cite{SmotherSpectrePortContention2019}. 
Also, similar to prior solutions~\cite{weisse2019nda, loughlin2021dolma, yu2019speculative} we prevent speculative store bypass attacks~\cite{horn2018speculative} by always sending requests for loads even if they match with an older, unresolved store (Load/Store Queue is hashed in Figure~\ref{fig:design}).

\section{Security Analysis}
\label{sec:sec-analysis}

A processor with secure speculation guarantees the absence of additional vulnerabilities on top of the non-speculative leaks of a given program (\textit{i.e.}, the leaks during the correct execution).
Two main programming policies (also known as \textit{software contracts}~\cite{guarnieri2021hardware}) 
are (1) \textit{sandboxing}, and (2) \textit{constant-time} execution that a processor requires
to protect them with secure speculation.
These policies are defined as followed:

\begin{definition}[Sandboxing policy]
\label{def:sb}
The sandboxing policy requires the program to ensure that all memory accesses are not outside the authorized address range.
\end{definition}

\begin{definition}[Constant-time policy]
\label{def:ct}
The constant-time policy requires that all non-speculative observations of the program (\textit{e.g.}, program counter and memory addresses) are independent and not affected by
secret values.
\end{definition}

\textbf{\name{} secure speculation for sandboxing and constant time policies}: 
To provide secure speculation for the sandboxing policy, the hardware
should restrict the execution of speculative memory accesses and their dependent instructions.
However, secure speculation for constant-time programs (commonly used for cryptographic algorithms) requires the hardware to avoid speculative execution of the instructions that are tainted by secret values, even if the secret itself has been loaded non-speculatively. \name{} supports secure speculation for both sandboxed and constant-time programs since it restricts the execution of \textit{all} speculative instructions. The performance benefits of \name{} come from allowing the execution of the instructions that are guaranteed to be independent of the speculation sources and speculative instructions. 
In other words, \name{} assumes that all instructions after speculation sources (\textit{e.g.}, conditional branches and indirect jumps) are secret dependent, unless the compiler declassifies the instructions that are independent of the speculation sources (\textit{i.e.}, they will leak during the non-speculative execution as well).
While our static compiler analysis 
might over-approximate the dependencies, it
is sound by design and will not declassify any true branch dependency. 

\begin{table}[t]
\centering
\scalebox{0.8}{
\begin{tabular}{c|l}
\toprule
\textbf{Design} & \multicolumn{1}{c}{\textbf{Description}}\\
\midrule
Unprotected & \multirow{2}{*}{Normal out-of-order core with no protection}\\
Baseline & \\
\hline
\name{} & Our proposed design\\
\hline
\multirow{2}{*}{STT} & STT and DOLMA implementations adopted\\
& from~\cite{loughlin2021dolma} to support secure speculation for\\
\cline{1-1}
\multirow{3}{*}{DOLMA} & sandboxing and constant-time\tablefootnote{We disable the delay-on-miss optimization of DOLMA as it is shown to be vulnerable to speculative interference attacks~\cite{behnia2021speculative}.} (as defined\\
& STT-Spectre (M+R) and  DOLMA-Default (M+R))\\
\hline
Secure Baseline & Disabling speculation (no branch predictor)\\
\bottomrule
\end{tabular}
}
\caption{Evaluated designs in our studies.}
\label{tb:designs}
\end{table}

In addition to previous software contracts, we introduce a new policy, called \textit{control-flow confidentiality}:

\begin{definition}[control-flow confidentiality policy]
\label{def:cfp}
Control-flow confidentiality policy requires the program to mark confidential control-flow instructions and guarantee that the correct execution of the program will not leak confidential information based on the adversaries' capabilities.
\end{definition}

\textbf{\name{} secure speculation for control-flow confidentiality}: 
To provide such a guarantee, \name{}
avoids any speculative activity (including the front-end) for confidential control-flow instructions. \name{} is the first solution that supports secure speculation for control-flow confidentiality through its software interface. 
\name{} provides the liberty for the software-level defenses to explore lightweight mitigations for non-speculative leaks based on the adversaries' capabilities without additional vulnerabilities due to the microarchitectural level speculations.

\section{Evaluation}
\label{sec:evaluation}

We evaluate \name{} compared to an unprotected baseline out-of-order core, and a secure baseline with the same protection scope as \name{} (addressing both speculative fetch and execution attacks). Additionally, we compare the modified version of DOLMA~\cite{loughlin2021dolma} and STT~\cite{yu2019speculative} which are still vulnerable to speculative fetch attacks.
Table~\ref{tb:designs} shows more details of our evaluated designs\footnote{
Note, that the STT and DOLMA designs referred to in the Related Work section and Table~\ref{tb:related-work} are the original designs, and are different from the versions we use for the evaluation.}.

\subsection{Experimental Setup}
\label{sec:setup}

\textbf{Simulation Environment}.
We implemented \name{} on top of the gem5~\cite{binkert2011gem5} DerivO3CPU and run simulations in syscall emulation (SE) mode.
For power analysis, we modify McPAT~\cite{li2013mcpat} version 1.3 to support our microarchitecture. We used a Golden-Cove-like microarchitecture~\cite{golden-cove} used in Intel Alder Lake processors as our baseline processor design (see Table~\ref{tb:setup} for the details of gem5 configuration). 

\begin{table}[t]
  \centering
  \resizebox{0.99\linewidth}{!}{%
  \begin{tabular}{@{}l|l||l|l@{}}
    \toprule
    L1d Cache & 32KB, 8-way & Branch Predictor & TAGE-SC-L-64KB\\
    \hline
    L1i Cache & 32KB, 8-way & LQ/SQ size & 192/114 entries\\
    \hline
    L2 Cache & 256KB, 8-way & ROB size & 512 entries\\
    \hline
    L3 Cache & 1MB, 16-way & IQ size & 97 entries\\
    \hline
    F/D/I/C width & 8/8/8/8 & RF (INT/FP) size & 280/332 entries\\
    \hline
    Data Prefetcher & Stride & UBT size & 16 entries\\
    \bottomrule
  \end{tabular}}
  \caption{System configuration for simulation (Intel Golden-Cove-like processor). 
  }
  \label{tb:setup}
\end{table}

\begin{figure}[t]
    \centering
    \includegraphics[trim=0 1.5cm 0 1cm,width=\linewidth]{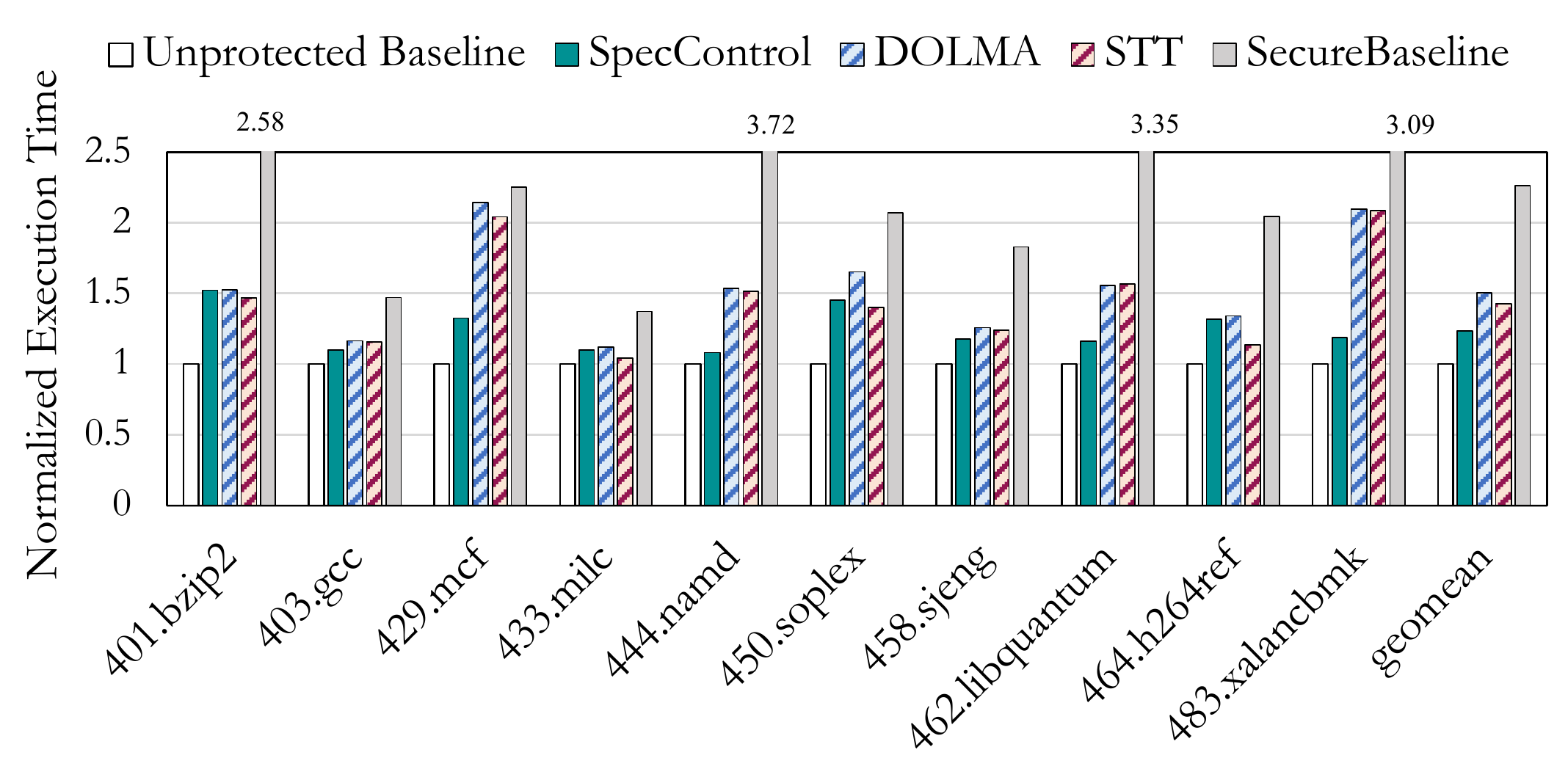}
    \caption{Performance results of SPEC CPU2006~\cite{henning2006spec}. \name{} protects all known speculative fetch and speculative execution attacks, while DOLMA and STT (hashed bars) only protect against speculative execution attacks. 
    }
    \label{fig:spec-perf}
\end{figure}

\textbf{Compiler Implementation}.
We implemented our compiler pass in LLVM 10.0~\cite{lattner2004llvm} to detect and mark branch dependencies. 
Our pass is built at the machine level for x86 architecture, but it is not architecture-specific and can be easily ported to other architectures.
In addition, we integrated a taint tracking mechanism in our compiler to mark secret dependent branches for crypto modules~\cite{borrello2021constantine}.

\textbf{Benchmarks}. We use C/C++ applications from SPEC CPU2006~\cite{henning2006spec} benchmark suite
and the ELFie~\cite{patil2021elfies} methodology to generate representative (SimPoint) executables 
with a region size of 1 billion instructions.
Also, we investigate a set of realistic crypto modules and standard microbenchmarks for constant-time enforcements~\cite{borrello2021constantine,rane2015raccoon,liu2015ghostrider,wu2018eliminating}.

\subsection{Performance Results}

\subsubsection{Performance of SPEC CPU2006 Workloads}
\label{sec:spec-perf}

Figure~\ref{fig:spec-perf} shows the performance results for SPEC CPU2006 applications for five different designs.
DOLMA and STT show an average performance overhead of \DOLMAPerfOverhead{} and \sttPerfOverhead{} compared to the Unprotected Baseline, respectively. 
\name{}'s performance overhead is just \SpecControlOverhead{} on average. This means that the \name{} methodology reduces the performance loss by \DOLMAPerfImprov{} over DOLMA and \STTPerfImprov{} over STT designs on average. 
In some cases STT shows better performance compared to \name{} (\textit{e.g.}, \texttt{464.h264ref}). The reason is that STT assumes that secret transmissions only happen through loads and does not restrict the execution of tainted stores, hence, restricting fewer instructions compared to \name{} and DOLMA in some cases.
However, prior work~\cite{loughlin2021dolma} demonstrates the vulnerability of STT against Spectre attacks transmitting data via stores. 
Note, that none of the branches in SPEC CPU2006 applications are marked as \resFE{} since they do not process confidential information.

Using a naive implementation (Secure Baseline) to protect the system against all speculation-based attacks will incur a significant performance degradation of \SBPerfOverhead{} on average compared to the Unprotected Baseline. 
\name{} provides the same protections while
showing only \SpecControlOverhead{} overhead.

\begin{figure}[t]
    \centering
    \includegraphics[trim=0 1.5cm 0 1cm,width=\linewidth]{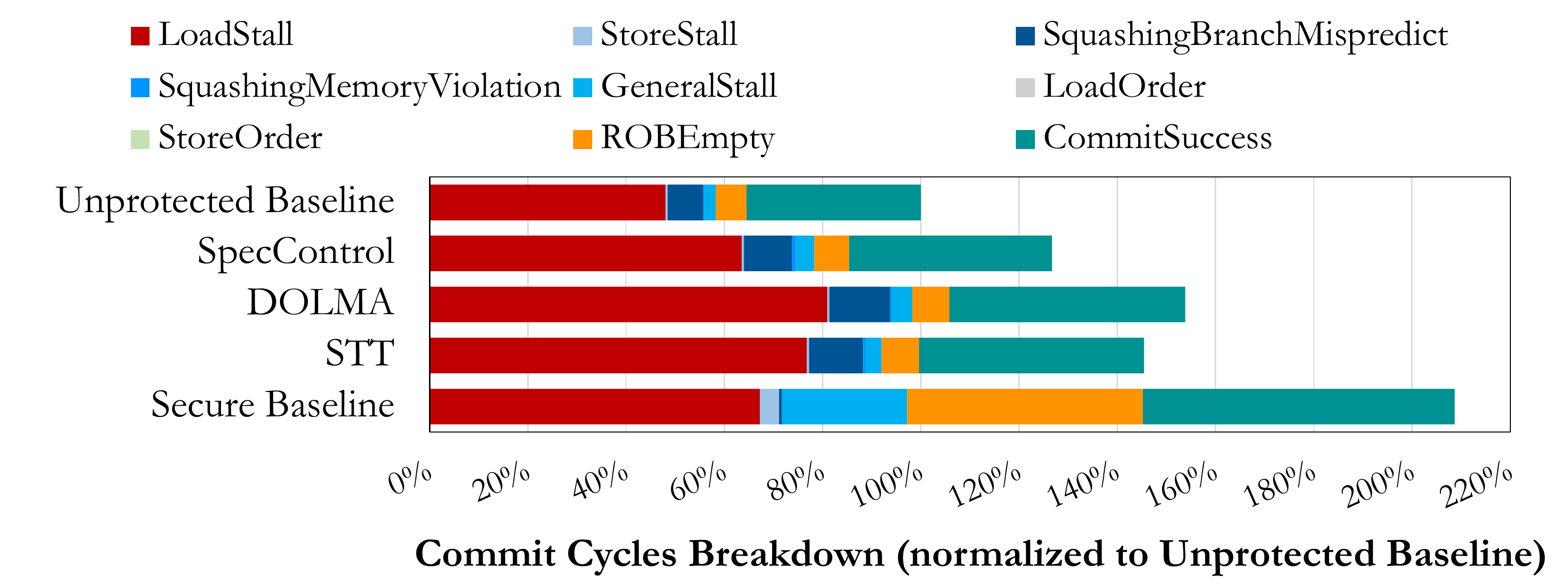}
    \caption{Commit cycle breakdown for different designs. Bars are normalized to the total commit cycles of the Unprotected Baseline. The LoadStall is the main reason for blocking the head of ROB.}
    \label{fig:commitCycles}
\end{figure}

\begin{figure}[t]
    \centering
    \includegraphics[trim=0 1.5cm 0 1cm,width=0.9\linewidth]{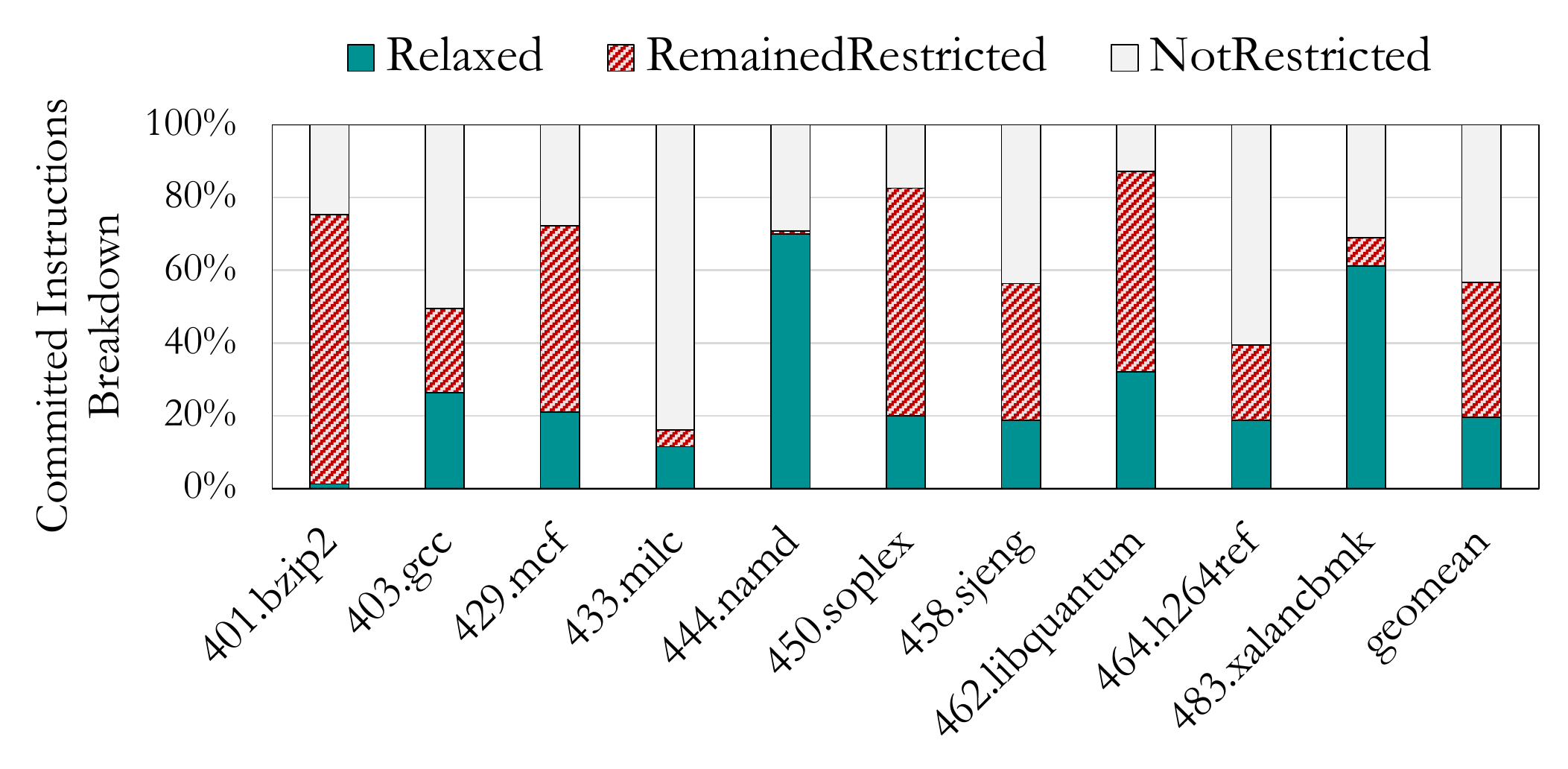}
    \caption{Committed instructions breakdown in \name{} based on their restrictions: (1) Relaxed: \name{} relaxed the restrictions based on compiler information, (2) RemainedRestricted: remained restricted until they resolved, (3) NotRestricted: the remaining instructions that executed normally. Note, that the instruction counts do not equate to performance.}
    \label{fig:instsBreakdown}
\end{figure}

Figure~\ref{fig:commitCycles} depicts the commit cycles breakdown (\textit{i.e.}, the cycles that at least on instruction is committed) of different designs (bars are normalized to the total commit cycles of the Unprotected Baseline). \textit{LoadStall} is the main reason for blocking the head of the ROB as a result of speculation restrictions. This occurs because restricting the execution of instructions will delay memory requests and lead to more cycles waiting for the data. 
\name{} is able to bring down the 81\% and 77\% of LoadStall in DOLMA and STT to 63\%, due to relaxing unnecessary restrictions informed by the compiler.
Figure~\ref{fig:instsBreakdown} shows the percentage of committed instructions in \name{} for different applications that were initially restricted (dependent on an unresolved branch) but \name{} was able to relax the restriction later (when the true dependent branches resolved). %
\name{} relaxes the restriction of 20\% of committed instructions on average based on the compiler information.
Some applications benefit from the compiler information significantly. For example, in \texttt{483.xalancbmk} 61\% of the restricted instructions are relaxed and \name{} shows only \xalancSpecControlOverhead{} performance overhead, compared to \xalancDOLMASTTOverhead{} overhead in DOLMA and STT.

\subsubsection{Impacts of UBT Size}
\label{sec:ubt-study}

\begin{figure}[t]
    \centering
    \includegraphics[trim=0 0.9cm 0 0.5cm,width=0.95\linewidth]{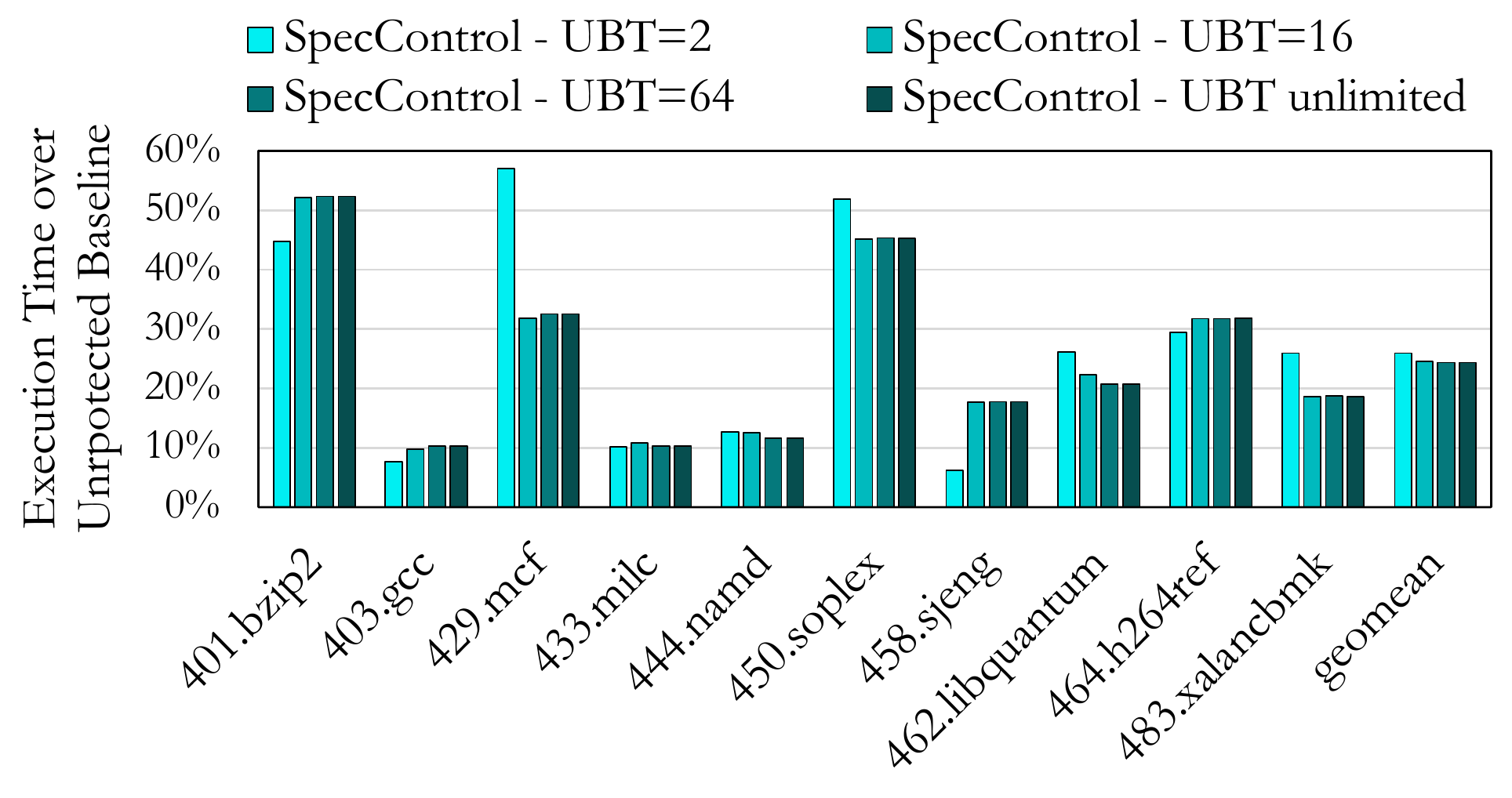}
    \caption{Execution time of \name{} over the Unprotected Baseline with different UBT sizes.}
    \label{fig:perf-ubt}
\end{figure}

\begin{figure}[t]
    \centering
    \includegraphics[trim=0 1cm 0 0,width=0.9\linewidth]{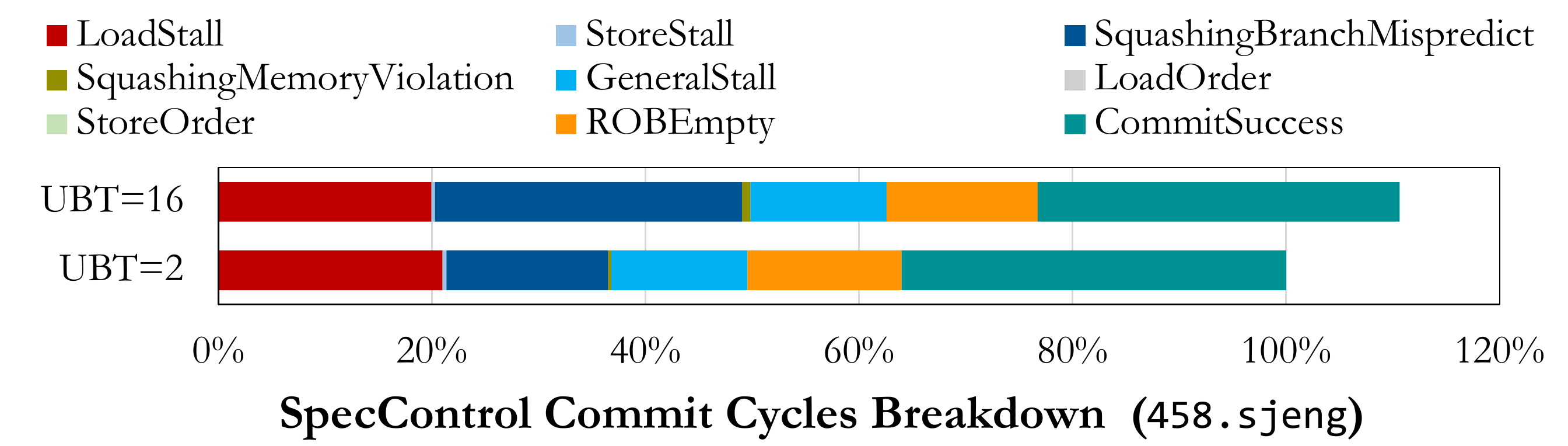}
    \caption{Breakdown of \name{} commit cycles for \texttt{458.sjeng} with UBT sizes of 2 and 16 (normalized to UBT=2). The main difference is the number of cycles spent on squashing due to branch misprediction.}
    \label{fig:sjeng-breakdown}
\end{figure}

\begin{figure*}[t]
    \centering
    \includegraphics[trim=0 1cm 0 1cm,width=0.95\linewidth]{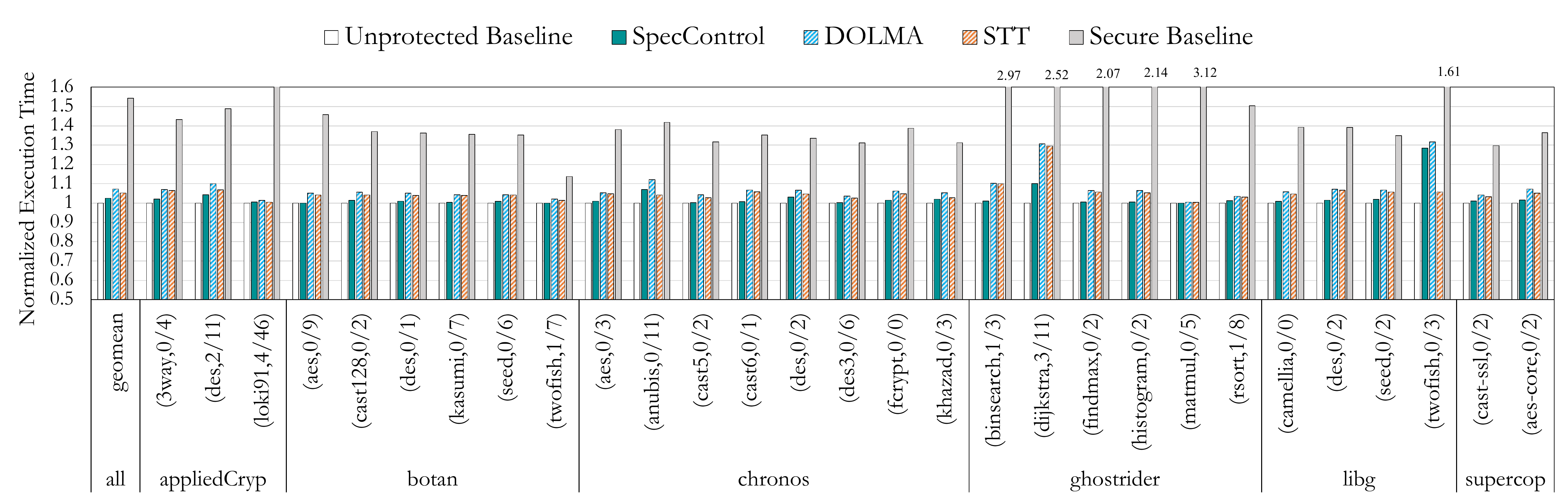}
    \caption{Performance results of cryptographic workloads. \name{} protects all known speculative fetch and speculative execution attacks, while DOLMA and STT (hashed bars) only protect speculative execution attacks. The x-axis shows the number of branches and sensitive (\textit{i.e.}, secret dependent) branches for each application (\textit{sensitive/total}).}
    \label{fig:perf-const}
\end{figure*}

Figure~\ref{fig:perf-ubt} shows the performance impacts for different UBT sizes. The expectation is higher overhead for smaller UBT sizes since it will block inserting instructions to the ROB more frequently (like \texttt{429.mcf}, and \texttt{450.soplex}). However, we observe that in some cases the \name{} overhead decreases when using a smaller UBT (\textit{e.g.}, \texttt{458.sjeng}).
More in-depth investigation shows that limiting the number of unresolved branches in some applications reduces the number of squashing cycles and the number of dynamic instructions that the core decodes and re-executes. 
Figure~\ref{fig:sjeng-breakdown} depicts the breakdown of the commit cycles for \texttt{458.sjeng} with UBT sizes of 2 and 16. The number of squashing cycles due to branch misprediction reduces by $1.9 \times$ for a UBT size of 2 compared to a UBT size of 16.
A potential future work can deploy a performance-aware and dynamic control of speculation level to limit the squashing cycles for problematic branches.
To have a balanced trade-off for the performance, power, and area, we use a UBT size of 16 entries (as a direct-mapped memory).
\name{} consumes only 1.72\% more power over the Unprotected Baseline core with an area overhead of 1.42\%.

\subsubsection{Performance of Cryptographic Workloads}

Figure~\ref{fig:perf-const} shows the performance results for our cryptographic benchmarks. We mark all sensitive branches of these applications via secret taint tracking and restrict them at the front-end (\resFE{}).
These applications are much smaller than SPEC CPU2006 with less intense control-flow leading to fewer opportunities to improve performance.
However, the results show that \name{} is still able to benefit from the compiler informed branch dependencies and improve the performance of these applications compared to STT and DOLMA, which only mitigate speculative execution attacks. For example, \name{} shows 10\% performance overhead for \texttt{dijkstra}, while STT and DOLMA incur 30\% overhead. For the same reason we explained for SPEC CPU2006, STT shows better performance for two applications as it does not restrict the execution of tainted stores. 

\subsubsection{Performance of Synthetic Workloads}

\begin{figure}[t]
    \centering
    \includegraphics[trim=0 0.5cm 0 0.5cm,width=0.8\linewidth]{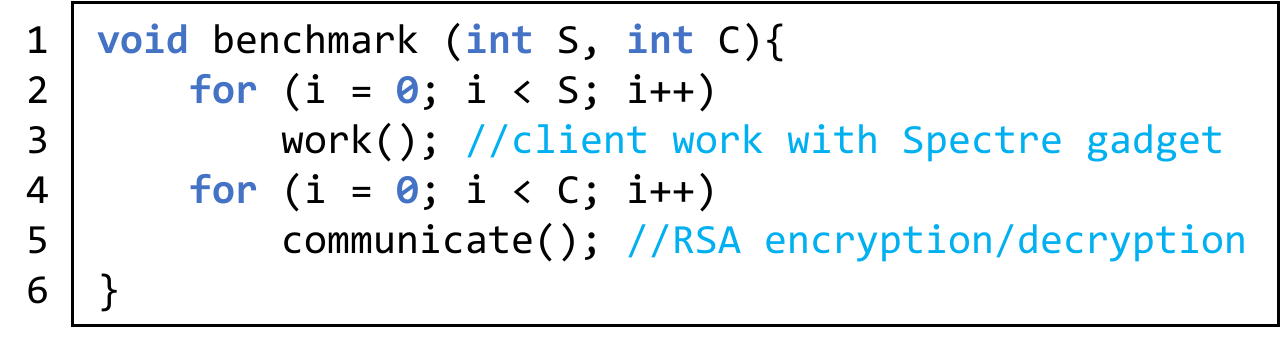}
    \caption{Pseudo code of our synthetic benchmark; \texttt{S} and \texttt{C} parameters specify the workloads' configuration.}
    \label{fig:syn-code}
\end{figure}

To better understand the effects of different configurations of \name{} in a real scenario,
we analyze a synthetic workload (Figure~\ref{fig:syn-code}) that mimics the behavior of a runtime application (\textit{e.g.}, JavaScript engine in a browser) that runs a (\texttt{S})andboxed application as a client workload and a (\texttt{C})rypto module for background communications~\cite{fustos2019spectreguard,daniel2023prospect}. \texttt{work()} contains sandboxes and Spectre gadgets, and \texttt{communicate()} is a Montgomery Ladder implementation of RSA.
The \texttt{S} and \texttt{C} parameters specify the size of each workload.

\begin{figure}[t]
    \centering
    \includegraphics[trim=4cm 1cm 4cm 1cm,width=0.75\linewidth]{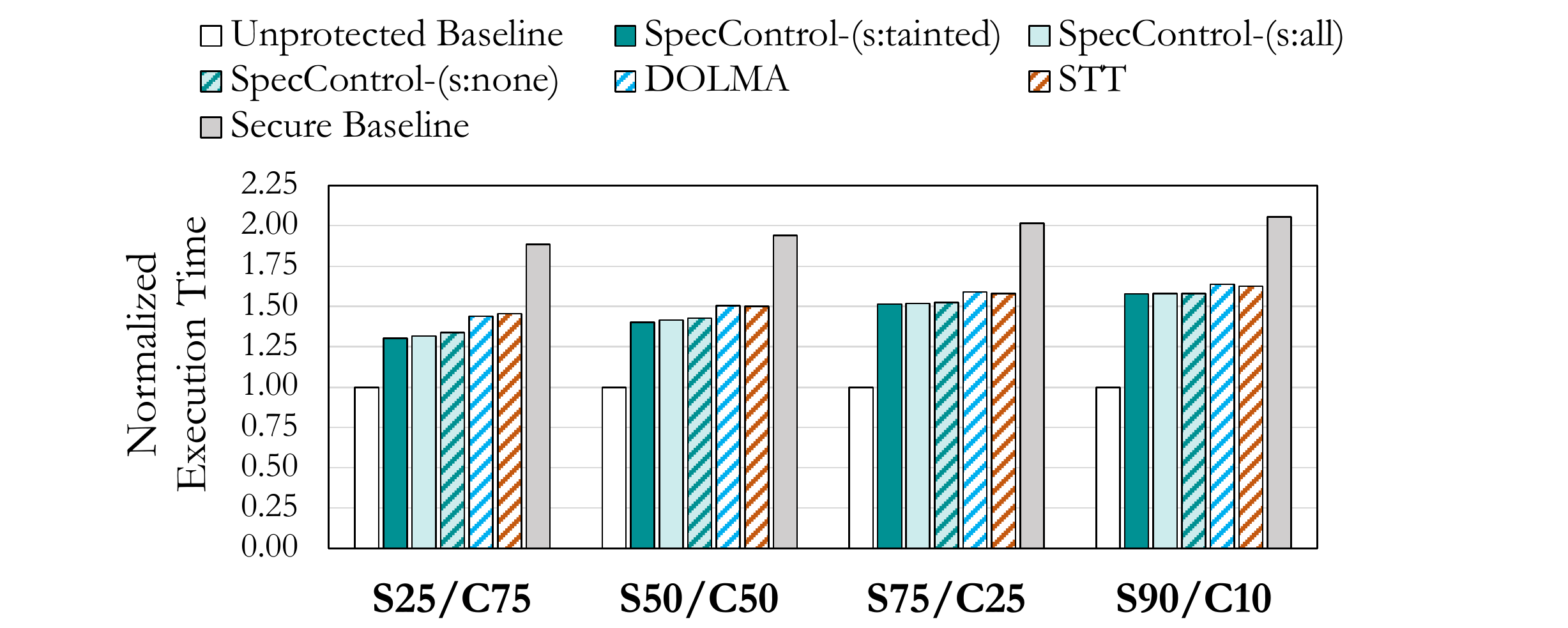}
    \caption{Performance of different synthetic workloads. \name{}-(s:tainted) marks the secret dependent branch in RSA as \resFE{}, \name{}-(s:all) marks all branches in RSA-related functions, and \name{}-(s:none) marks no branches (vulnerable to speculative fetch attacks).}
    \label{fig:syn-perf}
\end{figure}

Figure~\ref{fig:syn-perf} depicts the performance of different synthetic workloads. We evaluate three versions of \name{}: (1) \name{}-(s:tainted) restricts only the secret dependent branch of RSA at the front-end (line 4 in Figure~\ref{fig:RSA-decode}), (2) \name{}-(s:all) adopts a more coarse-grained approach and marks all branches in RSA-related functions, and (3) \name{}-(s:none) that adds no front-end restriction and have the same protection of DOLMA and STT.
As you can see, by decreasing the size of (\texttt{C})rypto part (as a regular application with less benefit from speculative execution) and increasing the size of (\texttt{S})andboxed part (more benefits from speculation) the performance overhead of \name{} increases, from 30\% in S25/C75 to 57\% in S90/C10 (STT and DOLMA show the same trend, from 45\% overhead to 64\%). In addition, for RSA we cannot observe a significant improvement by precisely marking front-end restrictions and a conservative approach is also sufficient to gain the most benefits of \name{}.
Interestingly, we observe that \name{}-(s:none) shows a bit more performance loss compared to \name{}-(s:tainted) (about 4\% for S25/C75 workload). This means that not using branch prediction for the tainted branch improves the performance and our simulation results show that the number of squashed instructions drop by 36\% compared to the Unprotected Baseline. 

\begin{figure}[t]
    \centering
    \includegraphics[trim=0 0.8cm 0 0,width=0.95\linewidth]{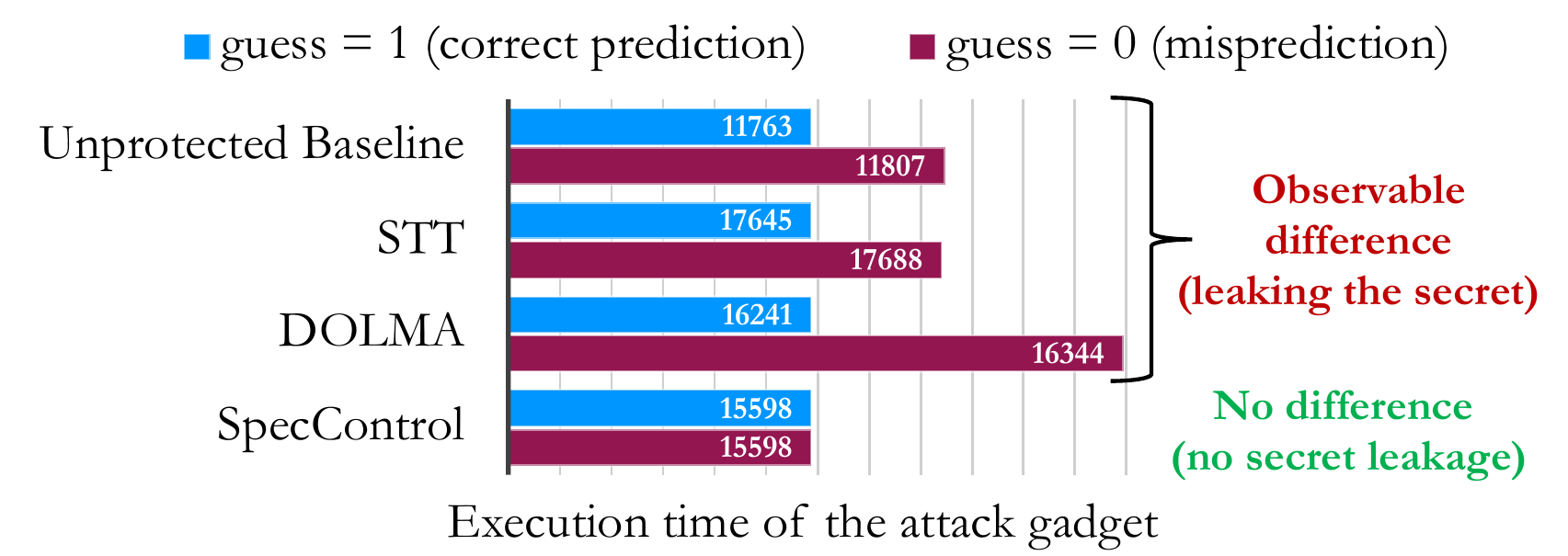}
    \caption{Speculative fetch attack results on gem5. 
    }
    \label{fig:gem5-attack}
\end{figure}

\begin{table*}[t]
\centering
\scalebox{0.85}{
\begin{tabular}{l|c|c|c}
\toprule
& \textbf{Intel} & \textbf{AMD} & \textbf{Apple} \\
\midrule
Model Name       & Intel(R) Core(TM) i7-4770 CPU @ 3.40GHz & AMD EPYC 7532 32-Core Processor & MacBook Pro (Apple M1 Pro chip)\\ \hline
CPU Cores        & 4     & 32 & 10 (8 performance and 2 efficiency) \\ 
\hline
CPU MHz          & 849.410 & 1499.479 & 2064 - 3220\\ 
\hline
Cache Size       & 8192KB & 512KB & L1: 128KB (perf.) 64KB (eff.), L2: 4MB \\ 
\hline
Operating System & Ubuntu 18.04 & Ubuntu 20.04 & macOS Ventura 13.0 \\ 
\hline
Compiler         & gcc (Ubuntu 7.5.0-3ubuntu1~18.04) 7.5.0, -O0 & GCC 9.4.0, -O0 & Apple clang version 14.0.0, -O0\\ 
\bottomrule
\end{tabular}
}
\caption{Experimental setup for speculative fetch attack on three different CPUs.}
\label{tb:real-attacks-setup}
\end{table*}

\begin{table}[t]
\centering
\scalebox{0.8}{
\begin{tabular}{c||r|r|r}
\toprule
\multirow{2}{*}{CPU} & \multicolumn{2}{c|}{Average Timing for} & \multirow{2}{*}{Min. Difference}\\
& Correct Guess & Incorrect Guess & \\
\midrule
Intel & 2726.65 & 6233.10 & 1657\\ 
\hline
AMD &  2101.20 & 2478.00 & 24\\
\hline
Apple\tablefootnote{Since there is no native implementation of \texttt{\_rdtsc()} on Apple processors, we implemented 
a timer function by incrementing a shared variable with the \texttt{OSAtomicIncrement64()} function.} & 185.45 & 823.30 & 23 \\
\bottomrule
\end{tabular}}
\caption{Speculative fetch attack results on real CPUs. 
}
\label{tb:real-attacks}
\end{table}

\subsection{Penetration Tests}

Figure~\ref{fig:gem5-attack} shows the penetration tests of our new variant of speculative fetch attacks on gem5. The figure shows the execution time of the attack gadget when the attacker chooses a correct or incorrect guess. A higher timing difference for the incorrect guess means that the attacker can recover the secret.
The results confirm that all other designs (the Unprotected Baseline, DOLMA, and STT) leak the secret bit through an observable timing difference, while \name{} can successfully protect the secret bit by restricting the fetch and front-end updates for the marked branch. 

In addition, we confirm that \name{} passes the Spectre penetration tests for both speculative secrets (sandboxing case) and non-speculative secrets (constant-time case)\footnote{We used the gem5 attacks provided by \cite{loughlin2021dolma}.}.

\subsection{Speculative Fetch Attack on existing CPUs}
\label{sec:real-attacks}

To demonstrate the feasibility of the attack on real existing CPUs, we run our attack on three different real CPUs, specified in Table~\ref{tb:real-attacks-setup}: (1) Intel, (2) AMD, and (3) Apple.
Table~\ref{tb:real-attacks} shows the average execution time of the attack gadget (running 10K NOPs) for 20 trials on these three CPUs with both correct and incorrect guesses. 
We observe that the execution time in the case of an incorrect guess is always higher than the case the attacker has a correct guess. For example for the Intel CPU, we observe at least 1657 cycles of timing difference for incorrect guesses and correct guesses which can be used as a threshold to evaluate the correctness of the guess.

\section{Related Work}
\label{sec:related_work}

\textbf{Channel-specific mitigations}.
Early defenses for speculative execution attacks aimed to secure individual channels, like data caches~\cite{Bourgeat2019MI6,khasawneh2019safespec,qureshi2018ceaser,saileshwar2019cleanupspec,DOM,yan2018invisispec}. Most of these solutions either implement invisible speculation of memory accesses or undo the speculation effects on the cache. Both of these two strategies are shown to be ineffective against newer versions of speculative execution attacks~\cite{behnia2021speculative,li2022unxpec}.
In addition, Weisse et al.~\cite{weisse2019nda} show that the attacker can exploit other channels as well to leak secret data, like BTB.
InvarSpec~\cite{zhao2020invarspec} is a performance optimization for invisible speculation techniques that uses program analysis to lift the restrictions for speculatively invariant instructions. Besides that InvarSpec inherits the security flaws of the underlying defenses, it also introduces new vulnerabilities~\cite{aimoniotis2021itsatrap}. 

\textbf{Secure speculation for sandboxing}. STT~\cite{yu2019speculative} and NDA \cite{weisse2019nda} are the first defenses that introduced taint tracking methodologies to restrict the execution of instructions tainted by speculative memory accesses. While STT restricts the \textit{execution} of tainted memory accesses, NDA prevents the \textit{propagation} of the data from tainted instructions. 
STT does not restrict the execution of tainted stores which has been shown to be vulnerable~\cite{loughlin2021dolma}.
Following mitigations adopted the same insights of speculative taint tracking while improving the performance with the same security guarantees~\cite{SDOSTT2020Yu,barber2019specshield}.
DOLMA~\cite{loughlin2021dolma} attempts to protect non-speculative secrets, but their performance benefits come from allowing the execution of some speculative instructions (under certain conditions) that might lead to exploits through resource contention~\cite{behnia2021speculative}.

\begin{table}[t]
\centering
\resizebox{\linewidth}{!}{%
\begin{threeparttable}[b]
\begin{tabular}{l|c|c|c|c|c}
\toprule
 & \multicolumn{3}{c|}{Secure Speculation} & & \\

Defense & \rotatebox[origin=c]{0}{\parbox{1.5cm}{\centering sandboxing}} & \rotatebox[origin=c]{0}{\parbox{1.5cm}{\centering constant-time}} & \rotatebox[origin=c]{0}{\parbox{2cm}{\centering control-flow\\confidentiality}} & \parbox{2cm}{\centering Security for\\legacy software$\ddagger$} & \parbox{2cm}{Flexibility for \\future policies}\\

\midrule
STT~\cite{yu2019speculative} & \halfcirc & \emptycirc & \emptycirc & \cmark & \xmark\\
\hline
NDA~\cite{weisse2019nda} & \fullcirc & \emptycirc & \emptycirc & \cmark & \xmark\\
\hline
DOLMA~\cite{loughlin2021dolma} & \halfcirc & \halfcirc & \emptycirc & \cmark & \xmark \\
\hline
ConTExT~\cite{schwarz2020context} & \fullcirc & \fullcirc & \emptycirc & \xmark & \xmark \\
\hline
ProSpeCT~\cite{daniel2023prospect} & \fullcirc & \fullcirc & \emptycirc & \xmark & \xmark \\
\hline
SPT~\cite{choudhary2021speculative} & \fullcirc & \fullcirc & \emptycirc & \cmark & \xmark \\
\hline
\rowcolor{JungleGreen!15}%
\name{} & \fullcirc & \fullcirc & \fullcirc & \cmark & \cmark\\
\bottomrule        
\end{tabular}
\begin{tablenotes}
    \centering
    \item \emptycirc~: no protection, \halfcirc~: partial protection, \fullcirc~: full protection
\end{tablenotes}
\end{threeparttable}
} %
\caption{Comparison of existing channel-agnostic methods for secure speculation. $\ddagger$ Regarding the provided protection scope (except for control-flow confidentiality).}
\label{tb:related-work}
\end{table}

\textbf{Secure speculation for constant-time}.
Some works~\cite{fustos2019spectreguard,schwarz2020context,daniel2023prospect} adopt a dynamic secrecy tracking methodology by manually labeling the secret regions of the memory; speculative execution of the instructions tainted by secret data will be blocked in the processor.
However, these defenses cannot provide security for legacy software without labeling the secret regions.
SPT~\cite{choudhary2021speculative} is a hardware-only defense supporting secure speculation for constant-time workloads that assumes all memory regions are secret and declassifies a region only if it is leaked non-speculatively as well.

Table~\ref{tb:related-work} shows a summary of existing defenses with respect to the secure speculations they offer. \name{} offers secure speculation for sandboxing and constant-time policies and is the only one to offer secure speculation for control-flow confidentiality. In addition, \name{} provides the flexibility to define new software contracts and restrict additional or fewer instructions through its interface based on future speculative attacks and programming policies.

\section{Conclusion}
\label{sec:conclusion}

In this work, we present \name{}, a configurable methodology that implements a hardware/software co-design to efficiently protect against current and future speculation-based attacks. \name{} proposes a communication interface that allows compilers and application developers to inform the hardware of the true branch dependencies, confidential instructions, and additional fine-grained constraints in order to apply restrictions only when necessary. \name{} minimizes performance overheads while allowing for configurable secure speculation policies. Apart from known speculative execution attacks, we also evaluate \name{} on a new variant of speculative fetch attacks that exploit the Pattern History Table (PHT) in branch predictors to extract the control-flow decisions of the victim. We show that this attack variant is more dangerous than previously reported. With \name{} we provide stronger security guarantees for speculation-based attacks 
compared to the state-of-the-art while reducing the performance overhead from \DOLMAPerfOverhead{} and \sttPerfOverhead{} to just \SpecControlOverhead{}.

\bibliographystyle{ACM-Reference-Format}
\bibliography{refs}


\begin{thebibliography}{78}


\ifx \showCODEN    \undefined \def \showCODEN     #1{\unskip}     \fi
\ifx \showDOI      \undefined \def \showDOI       #1{#1}\fi
\ifx \showISBNx    \undefined \def \showISBNx     #1{\unskip}     \fi
\ifx \showISBNxiii \undefined \def \showISBNxiii  #1{\unskip}     \fi
\ifx \showISSN     \undefined \def \showISSN      #1{\unskip}     \fi
\ifx \showLCCN     \undefined \def \showLCCN      #1{\unskip}     \fi
\ifx \shownote     \undefined \def \shownote      #1{#1}          \fi
\ifx \showarticletitle \undefined \def \showarticletitle #1{#1}   \fi
\ifx \showURL      \undefined \def \showURL       {\relax}        \fi
\providecommand\bibfield[2]{#2}
\providecommand\bibinfo[2]{#2}
\providecommand\natexlab[1]{#1}
\providecommand\showeprint[2][]{arXiv:#2}

\bibitem[mds({[n.\,d.]})]%
        {mds}
 \bibinfo{year}{[n.\,d.]}\natexlab{}.
\newblock \bibinfo{title}{Deep Dive: CPUID Enumeration and Architectural MSRs}.
\newblock
  \bibinfo{howpublished}{\url{https://www.intel.com/content/www/us/en/developer/topic-technology/software-security-guidance/overview.html\#MDS-CPUID}}.
\newblock


\bibitem[gnu({[n.\,d.]})]%
        {gnupg}
 \bibinfo{year}{[n.\,d.]}\natexlab{}.
\newblock \bibinfo{title}{GNU Privacy Guard}.
\newblock \bibinfo{howpublished}{\url{https://gnupg.org/}}.
\newblock


\bibitem[gol({[n.\,d.]})]%
        {golden-cove}
 \bibinfo{year}{[n.\,d.]}\natexlab{}.
\newblock \bibinfo{title}{Popping the Hood on Golden Cove}.
\newblock
  \bibinfo{howpublished}{\url{https://chipsandcheese.com/2021/12/02/popping-the-hood-on-golden-cove/}}.
\newblock


\bibitem[CBP(2016)]%
        {CBP}
 \bibinfo{year}{2016}\natexlab{}.
\newblock \showarticletitle{CBP-5 Kit}. In
  \bibinfo{booktitle}{\emph{Proceedings 5th Championship on Branch
  Prediction}}.
\newblock


\bibitem[bat(2021)]%
        {battery}
 \bibinfo{year}{2021}\natexlab{}.
\newblock \bibinfo{title}{Battery properties management}.
\newblock
  \bibinfo{howpublished}{\url{https://git.kernel.org/pub/scm/bluetooth/bluetooth-next.git/tree/drivers/hid/hid-input.c\#n247}}.
\newblock


\bibitem[blu(2021)]%
        {bluetooth}
 \bibinfo{year}{2021}\natexlab{}.
\newblock \bibinfo{title}{Bluetooth connection activity management}.
\newblock
  \bibinfo{howpublished}{\url{https://git.kernel.org/pub/scm/bluetooth/bluetooth-next.git/tree/drivers/hid/hid-input.c\#n383}}.
\newblock


\bibitem[Aimoniotis et~al\mbox{.}(2021)]%
        {aimoniotis2021itsatrap}
\bibfield{author}{\bibinfo{person}{Pavlos Aimoniotis},
  \bibinfo{person}{Christos Sakalis}, \bibinfo{person}{Magnus Själander},
  {and} \bibinfo{person}{Stefanos Kaxiras}.} \bibinfo{year}{2021}\natexlab{}.
\newblock \bibinfo{title}{"It's a Trap!"-How Speculation Invariance Can Be
  Abused with Forward Speculative Interference}.
\newblock
\newblock
\showeprint[arxiv]{2109.10774}~[cs.CR]


\bibitem[Barber et~al\mbox{.}(2019)]%
        {barber2019specshield}
\bibfield{author}{\bibinfo{person}{Kristin Barber}, \bibinfo{person}{Anys
  Bacha}, \bibinfo{person}{Li Zhou}, \bibinfo{person}{Yinqian Zhang}, {and}
  \bibinfo{person}{Radu Teodorescu}.} \bibinfo{year}{2019}\natexlab{}.
\newblock \showarticletitle{Specshield: Shielding speculative data from
  microarchitectural covert channels}. In \bibinfo{booktitle}{\emph{2019 28th
  International Conference on Parallel Architectures and Compilation Techniques
  (PACT)}}. IEEE, \bibinfo{pages}{151--164}.
\newblock


\bibitem[Barberis et~al\mbox{.}(2022)]%
        {barberis2022branch}
\bibfield{author}{\bibinfo{person}{Enrico Barberis}, \bibinfo{person}{Pietro
  Frigo}, \bibinfo{person}{Marius Muench}, \bibinfo{person}{Herbert Bos}, {and}
  \bibinfo{person}{Cristiano Giuffrida}.} \bibinfo{year}{2022}\natexlab{}.
\newblock \showarticletitle{Branch History Injection: On the Effectiveness of
  Hardware Mitigations Against $\{$Cross-Privilege$\}$ Spectre-v2 Attacks}. In
  \bibinfo{booktitle}{\emph{31st USENIX Security Symposium (USENIX Security
  22)}}. \bibinfo{pages}{971--988}.
\newblock


\bibitem[Behnia et~al\mbox{.}(2021)]%
        {behnia2021speculative}
\bibfield{author}{\bibinfo{person}{Mohammad Behnia}, \bibinfo{person}{Prateek
  Sahu}, \bibinfo{person}{Riccardo Paccagnella}, \bibinfo{person}{Jiyong Yu},
  \bibinfo{person}{Zirui~Neil Zhao}, \bibinfo{person}{Xiang Zou},
  \bibinfo{person}{Thomas Unterluggauer}, \bibinfo{person}{Josep Torrellas},
  \bibinfo{person}{Carlos Rozas}, \bibinfo{person}{Adam Morrison},
  {et~al\mbox{.}}} \bibinfo{year}{2021}\natexlab{}.
\newblock \showarticletitle{Speculative interference attacks: Breaking
  invisible speculation schemes}. In \bibinfo{booktitle}{\emph{Proceedings of
  the 26th ACM International Conference on Architectural Support for
  Programming Languages and Operating Systems}}. \bibinfo{pages}{1046--1060}.
\newblock


\bibitem[Bhattacharyya et~al\mbox{.}(2019)]%
        {SmotherSpectrePortContention2019}
\bibfield{author}{\bibinfo{person}{Atri Bhattacharyya},
  \bibinfo{person}{Alexandra Sandulescu}, \bibinfo{person}{Matthias
  Neugschwandtner}, \bibinfo{person}{Alessandro Sorniotti},
  \bibinfo{person}{Babak Falsafi}, \bibinfo{person}{Mathias Payer}, {and}
  \bibinfo{person}{Anil Kurmus}.} \bibinfo{year}{2019}\natexlab{}.
\newblock \showarticletitle{{SMoTherSpectre}}. \bibinfo{publisher}{{ACM}}.
\newblock
\urldef\tempurl%
\url{https://doi.org/10.1145/3319535.3363194}
\showDOI{\tempurl}


\bibitem[Binkert et~al\mbox{.}(2011)]%
        {binkert2011gem5}
\bibfield{author}{\bibinfo{person}{Nathan Binkert}, \bibinfo{person}{Bradford
  Beckmann}, \bibinfo{person}{Gabriel Black}, \bibinfo{person}{Steven~K
  Reinhardt}, \bibinfo{person}{Ali Saidi}, \bibinfo{person}{Arkaprava Basu},
  \bibinfo{person}{Joel Hestness}, \bibinfo{person}{Derek~R Hower},
  \bibinfo{person}{Tushar Krishna}, \bibinfo{person}{Somayeh Sardashti},
  {et~al\mbox{.}}} \bibinfo{year}{2011}\natexlab{}.
\newblock \showarticletitle{The gem5 simulator}.
\newblock \bibinfo{journal}{\emph{ACM SIGARCH Computer Architecture News}}
  \bibinfo{volume}{39}, \bibinfo{number}{2} (\bibinfo{year}{2011}),
  \bibinfo{pages}{1--7}.
\newblock


\bibitem[Borrello et~al\mbox{.}(2021)]%
        {borrello2021constantine}
\bibfield{author}{\bibinfo{person}{Pietro Borrello},
  \bibinfo{person}{Daniele~Cono D'Elia}, \bibinfo{person}{Leonardo Querzoni},
  {and} \bibinfo{person}{Cristiano Giuffrida}.}
  \bibinfo{year}{2021}\natexlab{}.
\newblock \showarticletitle{Constantine: Automatic Side-Channel Resistance
  Using Efficient Control and Data Flow Linearization}.
\newblock \bibinfo{journal}{\emph{arXiv preprint arXiv:2104.10749}}
  (\bibinfo{year}{2021}).
\newblock


\bibitem[Bourgeat et~al\mbox{.}(2019)]%
        {Bourgeat2019MI6}
\bibfield{author}{\bibinfo{person}{Thomas Bourgeat}, \bibinfo{person}{Ilia
  Lebedev}, \bibinfo{person}{Andrew Wright}, \bibinfo{person}{Sizhuo Zhang},
  \bibinfo{person}{Arvind}, {and} \bibinfo{person}{Srinivas Devadas}.}
  \bibinfo{year}{2019}\natexlab{}.
\newblock \showarticletitle{MI6: Secure Enclaves in a Speculative Out-of-Order
  Processor}. In \bibinfo{booktitle}{\emph{Proceedings of the 52nd Annual
  IEEE/ACM International Symposium on Microarchitecture}}
  \emph{(\bibinfo{series}{MICRO '52})}. \bibinfo{publisher}{Association for
  Computing Machinery}, \bibinfo{address}{New York, NY, USA},
  \bibinfo{pages}{42–56}.
\newblock
\showISBNx{9781450369381}
\urldef\tempurl%
\url{https://doi.org/10.1145/3352460.3358310}
\showDOI{\tempurl}


\bibitem[Canella et~al\mbox{.}(2019a)]%
        {canella2019fallout}
\bibfield{author}{\bibinfo{person}{Claudio Canella}, \bibinfo{person}{Daniel
  Genkin}, \bibinfo{person}{Lukas Giner}, \bibinfo{person}{Daniel Gruss},
  \bibinfo{person}{Moritz Lipp}, \bibinfo{person}{Marina Minkin},
  \bibinfo{person}{Daniel Moghimi}, \bibinfo{person}{Frank Piessens},
  \bibinfo{person}{Michael Schwarz}, \bibinfo{person}{Berk Sunar},
  {et~al\mbox{.}}} \bibinfo{year}{2019}\natexlab{a}.
\newblock \showarticletitle{Fallout: Leaking data on meltdown-resistant cpus}.
  In \bibinfo{booktitle}{\emph{Proceedings of the 2019 ACM SIGSAC Conference on
  Computer and Communications Security}}. \bibinfo{pages}{769--784}.
\newblock


\bibitem[Canella et~al\mbox{.}(2019b)]%
        {canella2019systematic}
\bibfield{author}{\bibinfo{person}{Claudio Canella}, \bibinfo{person}{Jo
  Van~Bulck}, \bibinfo{person}{Michael Schwarz}, \bibinfo{person}{Moritz Lipp},
  \bibinfo{person}{Benjamin Von~Berg}, \bibinfo{person}{Philipp Ortner},
  \bibinfo{person}{Frank Piessens}, \bibinfo{person}{Dmitry Evtyushkin}, {and}
  \bibinfo{person}{Daniel Gruss}.} \bibinfo{year}{2019}\natexlab{b}.
\newblock \showarticletitle{A systematic evaluation of transient execution
  attacks and defenses}. In \bibinfo{booktitle}{\emph{28th USENIX Security
  Symposium (USENIX Security 19)}}. \bibinfo{pages}{249--266}.
\newblock


\bibitem[Chandra et~al\mbox{.}(2001)]%
        {chandra2001parallel}
\bibfield{author}{\bibinfo{person}{Rohit Chandra}, \bibinfo{person}{Leo Dagum},
  \bibinfo{person}{David Kohr}, \bibinfo{person}{Ramesh Menon},
  \bibinfo{person}{Dror Maydan}, {and} \bibinfo{person}{Jeff McDonald}.}
  \bibinfo{year}{2001}\natexlab{}.
\newblock \bibinfo{booktitle}{\emph{Parallel programming in OpenMP}}.
\newblock \bibinfo{publisher}{Morgan kaufmann}.
\newblock


\bibitem[Chen et~al\mbox{.}(2019)]%
        {chen2019sgxpectre}
\bibfield{author}{\bibinfo{person}{Guoxing Chen}, \bibinfo{person}{Sanchuan
  Chen}, \bibinfo{person}{Yuan Xiao}, \bibinfo{person}{Yinqian Zhang},
  \bibinfo{person}{Zhiqiang Lin}, {and} \bibinfo{person}{Ten~H Lai}.}
  \bibinfo{year}{2019}\natexlab{}.
\newblock \showarticletitle{Sgxpectre: Stealing intel secrets from sgx enclaves
  via speculative execution}. In \bibinfo{booktitle}{\emph{2019 IEEE European
  Symposium on Security and Privacy (EuroS\&P)}}. IEEE,
  \bibinfo{pages}{142--157}.
\newblock


\bibitem[Chen et~al\mbox{.}(2023)]%
        {chen2023afterimage}
\bibfield{author}{\bibinfo{person}{Yun Chen}, \bibinfo{person}{Lingfeng Pei},
  {and} \bibinfo{person}{Trevor~E Carlson}.} \bibinfo{year}{2023}\natexlab{}.
\newblock \showarticletitle{AfterImage: Leaking Control Flow Data and Tracking
  Load Operations via the Hardware Prefetcher}. In
  \bibinfo{booktitle}{\emph{Proceedings of the 28th ACM International
  Conference on Architectural Support for Programming Languages and Operating
  Systems (ASPLOS), Volume 2}}. \bibinfo{pages}{16--32}.
\newblock


\bibitem[Choudhary et~al\mbox{.}(2021)]%
        {choudhary2021speculative}
\bibfield{author}{\bibinfo{person}{Rutvik Choudhary}, \bibinfo{person}{Jiyong
  Yu}, \bibinfo{person}{Christopher Fletcher}, {and} \bibinfo{person}{Adam
  Morrison}.} \bibinfo{year}{2021}\natexlab{}.
\newblock \showarticletitle{Speculative Privacy Tracking (SPT): Leaking
  Information From Speculative Execution Without Compromising Privacy}. In
  \bibinfo{booktitle}{\emph{MICRO-54: 54th Annual IEEE/ACM International
  Symposium on Microarchitecture}}. \bibinfo{pages}{607--622}.
\newblock


\bibitem[Daniel et~al\mbox{.}(2023)]%
        {daniel2023prospect}
\bibfield{author}{\bibinfo{person}{Lesly-Ann Daniel}, \bibinfo{person}{Marton
  Bognar}, \bibinfo{person}{Job Noorman}, \bibinfo{person}{S{\'e}bastien
  Bardin}, \bibinfo{person}{Tamara Rezk}, {and} \bibinfo{person}{Frank
  Piessens}.} \bibinfo{year}{2023}\natexlab{}.
\newblock \showarticletitle{ProSpeCT: Provably Secure Speculation for the
  Constant-Time Policy (Extended version)}.
\newblock \bibinfo{journal}{\emph{arXiv preprint arXiv:2302.12108}}
  (\bibinfo{year}{2023}).
\newblock


\bibitem[Evtyushkin et~al\mbox{.}(2018)]%
        {evtyushkin2018branchscope}
\bibfield{author}{\bibinfo{person}{Dmitry Evtyushkin}, \bibinfo{person}{Ryan
  Riley}, \bibinfo{person}{Nael~CSE Abu-Ghazaleh}, \bibinfo{person}{ECE}, {and}
  \bibinfo{person}{Dmitry Ponomarev}.} \bibinfo{year}{2018}\natexlab{}.
\newblock \showarticletitle{Branchscope: A new side-channel attack on
  directional branch predictor}.
\newblock \bibinfo{journal}{\emph{ACM SIGPLAN Notices}} \bibinfo{volume}{53},
  \bibinfo{number}{2} (\bibinfo{year}{2018}), \bibinfo{pages}{693--707}.
\newblock


\bibitem[Fan et~al\mbox{.}(2016)]%
        {fan2016attacking}
\bibfield{author}{\bibinfo{person}{Shuqin Fan}, \bibinfo{person}{Wenbo Wang},
  {and} \bibinfo{person}{Qingfeng Cheng}.} \bibinfo{year}{2016}\natexlab{}.
\newblock \showarticletitle{Attacking OpenSSL implementation of ECDSA with a
  few signatures}. In \bibinfo{booktitle}{\emph{Proceedings of the 2016 ACM
  SIGSAC Conference on Computer and Communications Security}}.
  \bibinfo{pages}{1505--1515}.
\newblock


\bibitem[Fustos et~al\mbox{.}(2019)]%
        {fustos2019spectreguard}
\bibfield{author}{\bibinfo{person}{Jacob Fustos}, \bibinfo{person}{Farzad
  Farshchi}, {and} \bibinfo{person}{Heechul Yun}.}
  \bibinfo{year}{2019}\natexlab{}.
\newblock \showarticletitle{Spectreguard: An efficient data-centric defense
  mechanism against spectre attacks}. In \bibinfo{booktitle}{\emph{Proceedings
  of the 56th Annual Design Automation Conference 2019}}.
  \bibinfo{pages}{1--6}.
\newblock


\bibitem[Gruss et~al\mbox{.}(2016)]%
        {gruss2016prefetch}
\bibfield{author}{\bibinfo{person}{Daniel Gruss},
  \bibinfo{person}{Cl{\'e}mentine Maurice}, \bibinfo{person}{Anders Fogh},
  \bibinfo{person}{Moritz Lipp}, {and} \bibinfo{person}{Stefan Mangard}.}
  \bibinfo{year}{2016}\natexlab{}.
\newblock \showarticletitle{Prefetch side-channel attacks: Bypassing SMAP and
  kernel ASLR}. In \bibinfo{booktitle}{\emph{Proceedings of the 2016 ACM SIGSAC
  conference on computer and communications security}}.
  \bibinfo{pages}{368--379}.
\newblock


\bibitem[Guarnieri et~al\mbox{.}(2021)]%
        {guarnieri2021hardware}
\bibfield{author}{\bibinfo{person}{Marco Guarnieri}, \bibinfo{person}{Boris
  K{\"o}pf}, \bibinfo{person}{Jan Reineke}, {and} \bibinfo{person}{Pepe Vila}.}
  \bibinfo{year}{2021}\natexlab{}.
\newblock \showarticletitle{Hardware-software contracts for secure
  speculation}. In \bibinfo{booktitle}{\emph{2021 IEEE Symposium on Security
  and Privacy (SP)}}. IEEE, \bibinfo{pages}{1868--1883}.
\newblock


\bibitem[Hajiabadi et~al\mbox{.}(2021)]%
        {hajiabadi2021noreba}
\bibfield{author}{\bibinfo{person}{Ali Hajiabadi}, \bibinfo{person}{Andreas
  Diavastos}, {and} \bibinfo{person}{Trevor~E Carlson}.}
  \bibinfo{year}{2021}\natexlab{}.
\newblock \showarticletitle{NOREBA: a compiler-informed non-speculative
  out-of-order commit processor}. In \bibinfo{booktitle}{\emph{Proceedings of
  the 26th ACM International Conference on Architectural Support for
  Programming Languages and Operating Systems}}. \bibinfo{pages}{182--193}.
\newblock


\bibitem[Henning(2006)]%
        {henning2006spec}
\bibfield{author}{\bibinfo{person}{John~L Henning}.}
  \bibinfo{year}{2006}\natexlab{}.
\newblock \showarticletitle{SPEC CPU2006 benchmark descriptions}.
\newblock \bibinfo{journal}{\emph{ACM SIGARCH Computer Architecture News}}
  \bibinfo{volume}{34}, \bibinfo{number}{4} (\bibinfo{year}{2006}),
  \bibinfo{pages}{1--17}.
\newblock


\bibitem[Horn(2018)]%
        {horn2018speculative}
\bibfield{author}{\bibinfo{person}{Jann Horn}.}
  \bibinfo{year}{2018}\natexlab{}.
\newblock \bibinfo{title}{speculative execution, variant 4: speculative store
  bypass}.
\newblock
\newblock


\bibitem[Jim{\'e}nez(2003)]%
        {jimenez2003fast}
\bibfield{author}{\bibinfo{person}{Daniel~A Jim{\'e}nez}.}
  \bibinfo{year}{2003}\natexlab{}.
\newblock \showarticletitle{Fast path-based neural branch prediction}. In
  \bibinfo{booktitle}{\emph{Proceedings. 36th Annual IEEE/ACM International
  Symposium on Microarchitecture, 2003. MICRO-36.}} IEEE,
  \bibinfo{pages}{243--252}.
\newblock


\bibitem[Jim{\'e}nez and Lin(2001)]%
        {jimenez2001dynamic}
\bibfield{author}{\bibinfo{person}{Daniel~A Jim{\'e}nez} {and}
  \bibinfo{person}{Calvin Lin}.} \bibinfo{year}{2001}\natexlab{}.
\newblock \showarticletitle{Dynamic branch prediction with perceptrons}. In
  \bibinfo{booktitle}{\emph{Proceedings HPCA Seventh International Symposium on
  High-Performance Computer Architecture}}. IEEE, \bibinfo{pages}{197--206}.
\newblock


\bibitem[Joye and Yen(2002)]%
        {joye2002montgomery}
\bibfield{author}{\bibinfo{person}{Marc Joye} {and} \bibinfo{person}{Sung-Ming
  Yen}.} \bibinfo{year}{2002}\natexlab{}.
\newblock \showarticletitle{The Montgomery powering ladder}. In
  \bibinfo{booktitle}{\emph{International workshop on cryptographic hardware
  and embedded systems}}. Springer, \bibinfo{pages}{291--302}.
\newblock


\bibitem[Khasawneh et~al\mbox{.}(2019)]%
        {khasawneh2019safespec}
\bibfield{author}{\bibinfo{person}{Khaled~N Khasawneh},
  \bibinfo{person}{Esmaeil~Mohammadian Koruyeh}, \bibinfo{person}{Chengyu
  Song}, \bibinfo{person}{Dmitry Evtyushkin}, \bibinfo{person}{Dmitry
  Ponomarev}, {and} \bibinfo{person}{Nael Abu-Ghazaleh}.}
  \bibinfo{year}{2019}\natexlab{}.
\newblock \showarticletitle{Safespec: Banishing the spectre of a meltdown with
  leakage-free speculation}. In \bibinfo{booktitle}{\emph{2019 56th ACM/IEEE
  Design Automation Conference (DAC)}}. IEEE, \bibinfo{pages}{1--6}.
\newblock


\bibitem[Kiriansky and Waldspurger(2018a)]%
        {kiriansky2018speculative}
\bibfield{author}{\bibinfo{person}{Vladimir Kiriansky} {and}
  \bibinfo{person}{Carl Waldspurger}.} \bibinfo{year}{2018}\natexlab{a}.
\newblock \showarticletitle{Speculative buffer overflows: Attacks and
  defenses}.
\newblock \bibinfo{journal}{\emph{arXiv preprint arXiv:1807.03757}}
  (\bibinfo{year}{2018}).
\newblock


\bibitem[Kiriansky and Waldspurger(2018b)]%
        {Kiriansky2018SBOSpectre}
\bibfield{author}{\bibinfo{person}{Vladimir Kiriansky} {and}
  \bibinfo{person}{Carl~A. Waldspurger}.} \bibinfo{year}{2018}\natexlab{b}.
\newblock \showarticletitle{Speculative Buffer Overflows: Attacks and
  Defenses}.
\newblock \bibinfo{journal}{\emph{CoRR}}  \bibinfo{volume}{abs/1807.03757}
  (\bibinfo{year}{2018}).
\newblock
\showeprint[arXiv]{1807.03757}
\urldef\tempurl%
\url{http://arxiv.org/abs/1807.03757}
\showURL{%
\tempurl}


\bibitem[Kirzner and Morrison(2021)]%
        {kirzner2021analysis}
\bibfield{author}{\bibinfo{person}{Ofek Kirzner} {and} \bibinfo{person}{Adam
  Morrison}.} \bibinfo{year}{2021}\natexlab{}.
\newblock \showarticletitle{An analysis of speculative type confusion
  vulnerabilities in the wild}. In \bibinfo{booktitle}{\emph{30th USENIX
  Security Symposium (USENIX Security 21)}}. \bibinfo{pages}{2399--2416}.
\newblock


\bibitem[Kocher et~al\mbox{.}(2019)]%
        {Spectre2019Kocher}
\bibfield{author}{\bibinfo{person}{Paul Kocher}, \bibinfo{person}{Jann Horn},
  \bibinfo{person}{Anders Fogh}, \bibinfo{person}{Daniel Genkin},
  \bibinfo{person}{Daniel Gruss}, \bibinfo{person}{Werner Haas},
  \bibinfo{person}{Mike Hamburg}, \bibinfo{person}{Moritz Lipp},
  \bibinfo{person}{Stefan Mangard}, \bibinfo{person}{Thomas Prescher},
  \bibinfo{person}{Michael Schwarz}, {and} \bibinfo{person}{Yuval Yarom}.}
  \bibinfo{year}{2019}\natexlab{}.
\newblock \showarticletitle{Spectre Attacks: Exploiting Speculative Execution}.
  \bibinfo{publisher}{{IEEE}}.
\newblock
\urldef\tempurl%
\url{https://doi.org/10.1109/sp.2019.00002}
\showDOI{\tempurl}


\bibitem[Koruyeh et~al\mbox{.}(2018a)]%
        {SpectreReturns2018Koruyeh}
\bibfield{author}{\bibinfo{person}{Esmaeil~Mohammadian Koruyeh},
  \bibinfo{person}{Khaled~N. Khasawneh}, \bibinfo{person}{Chengyu Song}, {and}
  \bibinfo{person}{Nael Abu-Ghazaleh}.} \bibinfo{year}{2018}\natexlab{a}.
\newblock \showarticletitle{Spectre Returns! Speculation Attacks using the
  Return Stack Buffer}. In \bibinfo{booktitle}{\emph{12th USENIX Workshop on
  Offensive Technologies (WOOT 18)}}. \bibinfo{publisher}{USENIX Association},
  \bibinfo{address}{Baltimore, MD}.
\newblock
\urldef\tempurl%
\url{https://www.usenix.org/conference/woot18/presentation/koruyeh}
\showURL{%
\tempurl}


\bibitem[Koruyeh et~al\mbox{.}(2018b)]%
        {koruyeh2018spectre}
\bibfield{author}{\bibinfo{person}{Esmaeil~Mohammadian Koruyeh},
  \bibinfo{person}{Khaled~N Khasawneh}, \bibinfo{person}{Chengyu Song}, {and}
  \bibinfo{person}{Nael~B Abu-Ghazaleh}.} \bibinfo{year}{2018}\natexlab{b}.
\newblock \showarticletitle{Spectre Returns! Speculation Attacks using the
  Return Stack Buffer.}. In \bibinfo{booktitle}{\emph{WOOT@ USENIX Security
  Symposium}}.
\newblock


\bibitem[Lattner and Adve(2004)]%
        {lattner2004llvm}
\bibfield{author}{\bibinfo{person}{Chris Lattner} {and} \bibinfo{person}{Vikram
  Adve}.} \bibinfo{year}{2004}\natexlab{}.
\newblock \showarticletitle{LLVM: A compilation framework for lifelong program
  analysis \& transformation}. In \bibinfo{booktitle}{\emph{International
  Symposium on Code Generation and Optimization (CGO '04)}}.
\newblock
\urldef\tempurl%
\url{https://doi.org/10.1109/CGO.2004.1281665}
\showDOI{\tempurl}


\bibitem[Lee et~al\mbox{.}(2017)]%
        {lee2017inferring}
\bibfield{author}{\bibinfo{person}{Sangho Lee}, \bibinfo{person}{Ming-Wei
  Shih}, \bibinfo{person}{Prasun Gera}, \bibinfo{person}{Taesoo Kim},
  \bibinfo{person}{Hyesoon Kim}, {and} \bibinfo{person}{Marcus Peinado}.}
  \bibinfo{year}{2017}\natexlab{}.
\newblock \showarticletitle{Inferring fine-grained control flow inside
  $\{$SGX$\}$ enclaves with branch shadowing}. In
  \bibinfo{booktitle}{\emph{26th USENIX Security Symposium (USENIX Security
  17)}}. \bibinfo{pages}{557--574}.
\newblock


\bibitem[Li et~al\mbox{.}(2022)]%
        {li2022unxpec}
\bibfield{author}{\bibinfo{person}{Mengming Li}, \bibinfo{person}{Chenlu Miao},
  \bibinfo{person}{Yilong Yang}, {and} \bibinfo{person}{Kai Bu}.}
  \bibinfo{year}{2022}\natexlab{}.
\newblock \showarticletitle{unXpec: Breaking Undo-based Safe Speculation}. In
  \bibinfo{booktitle}{\emph{2022 IEEE International Symposium on
  High-Performance Computer Architecture (HPCA)}}. IEEE,
  \bibinfo{pages}{98--112}.
\newblock


\bibitem[Li et~al\mbox{.}(2013)]%
        {li2013mcpat}
\bibfield{author}{\bibinfo{person}{Sheng Li}, \bibinfo{person}{Jung~Ho Ahn},
  \bibinfo{person}{Richard~D Strong}, \bibinfo{person}{Jay~B Brockman},
  \bibinfo{person}{Dean~M Tullsen}, {and} \bibinfo{person}{Norman~P Jouppi}.}
  \bibinfo{year}{2013}\natexlab{}.
\newblock \showarticletitle{The McPAT framework for multicore and manycore
  architectures: Simultaneously modeling power, area, and timing}.
\newblock \bibinfo{journal}{\emph{ACM Transactions on Architecture and Code
  Optimization (TACO)}} \bibinfo{volume}{10}, \bibinfo{number}{1}
  (\bibinfo{year}{2013}), \bibinfo{pages}{5}.
\newblock
\urldef\tempurl%
\url{https://doi.org/10.1145/2445572.2445577}
\showDOI{\tempurl}


\bibitem[Lipp et~al\mbox{.}(2022)]%
        {lipp2022amd}
\bibfield{author}{\bibinfo{person}{Moritz Lipp}, \bibinfo{person}{Daniel
  Gruss}, {and} \bibinfo{person}{Michael Schwarz}.}
  \bibinfo{year}{2022}\natexlab{}.
\newblock \showarticletitle{$\{$AMD$\}$ Prefetch Attacks through Power and
  Time}. In \bibinfo{booktitle}{\emph{31st USENIX Security Symposium (USENIX
  Security 22)}}. \bibinfo{pages}{643--660}.
\newblock


\bibitem[Lipp et~al\mbox{.}(2018)]%
        {Lipp2018meltdown}
\bibfield{author}{\bibinfo{person}{Moritz Lipp}, \bibinfo{person}{Michael
  Schwarz}, \bibinfo{person}{Daniel Gruss}, \bibinfo{person}{Thomas Prescher},
  \bibinfo{person}{Werner Haas}, \bibinfo{person}{Anders Fogh},
  \bibinfo{person}{Jann Horn}, \bibinfo{person}{Stefan Mangard},
  \bibinfo{person}{Paul Kocher}, \bibinfo{person}{Daniel Genkin},
  \bibinfo{person}{Yuval Yarom}, {and} \bibinfo{person}{Mike Hamburg}.}
  \bibinfo{year}{2018}\natexlab{}.
\newblock \showarticletitle{Meltdown: Reading Kernel Memory from User Space}.
  In \bibinfo{booktitle}{\emph{27th {USENIX} Security Symposium ({USENIX}
  Security 18)}}.
\newblock


\bibitem[Liu et~al\mbox{.}(2015)]%
        {liu2015ghostrider}
\bibfield{author}{\bibinfo{person}{Chang Liu}, \bibinfo{person}{Austin Harris},
  \bibinfo{person}{Martin Maas}, \bibinfo{person}{Michael Hicks},
  \bibinfo{person}{Mohit Tiwari}, {and} \bibinfo{person}{Elaine Shi}.}
  \bibinfo{year}{2015}\natexlab{}.
\newblock \showarticletitle{Ghostrider: A hardware-software system for memory
  trace oblivious computation}.
\newblock \bibinfo{journal}{\emph{ACM SIGPLAN Notices}} \bibinfo{volume}{50},
  \bibinfo{number}{4} (\bibinfo{year}{2015}), \bibinfo{pages}{87--101}.
\newblock


\bibitem[Loughlin et~al\mbox{.}(2021)]%
        {loughlin2021dolma}
\bibfield{author}{\bibinfo{person}{Kevin Loughlin}, \bibinfo{person}{Ian Neal},
  \bibinfo{person}{Jiacheng Ma}, \bibinfo{person}{Elisa Tsai},
  \bibinfo{person}{Ofir Weisse}, \bibinfo{person}{Satish Narayanasamy}, {and}
  \bibinfo{person}{Baris Kasikci}.} \bibinfo{year}{2021}\natexlab{}.
\newblock \showarticletitle{$\{$DOLMA$\}$: Securing Speculation with the
  Principle of Transient $\{$Non-Observability$\}$}. In
  \bibinfo{booktitle}{\emph{30th USENIX Security Symposium (USENIX Security
  21)}}. \bibinfo{pages}{1397--1414}.
\newblock


\bibitem[Mahlke et~al\mbox{.}(1992)]%
        {mahlke1992effective}
\bibfield{author}{\bibinfo{person}{Scott~A Mahlke}, \bibinfo{person}{David~C
  Lin}, \bibinfo{person}{William~Y Chen}, \bibinfo{person}{Richard~E Hank},
  {and} \bibinfo{person}{Roger~A Bringmann}.} \bibinfo{year}{1992}\natexlab{}.
\newblock \showarticletitle{Effective compiler support for predicated execution
  using the hyperblock}.
\newblock \bibinfo{journal}{\emph{ACM SIGMICRO Newsletter}}
  \bibinfo{volume}{23}, \bibinfo{number}{1-2} (\bibinfo{year}{1992}),
  \bibinfo{pages}{45--54}.
\newblock


\bibitem[Maisuradze and Rossow(2018)]%
        {maisuradze2018ret2spec}
\bibfield{author}{\bibinfo{person}{Giorgi Maisuradze} {and}
  \bibinfo{person}{Christian Rossow}.} \bibinfo{year}{2018}\natexlab{}.
\newblock \showarticletitle{ret2spec: Speculative execution using return stack
  buffers}. In \bibinfo{booktitle}{\emph{Proceedings of the 2018 ACM SIGSAC
  Conference on Computer and Communications Security}}.
  \bibinfo{pages}{2109--2122}.
\newblock


\bibitem[Mambretti et~al\mbox{.}(2019)]%
        {SpectreVariantCFH2019Mambretti}
\bibfield{author}{\bibinfo{person}{Andrea Mambretti},
  \bibinfo{person}{Alexandra Sandulescu}, \bibinfo{person}{Matthias
  Neugschwandtner}, \bibinfo{person}{Alessandro Sorniotti}, {and}
  \bibinfo{person}{Anil Kurmus}.} \bibinfo{year}{2019}\natexlab{}.
\newblock \showarticletitle{Two methods for exploiting speculative control flow
  hijacks}. In \bibinfo{booktitle}{\emph{13th USENIX Workshop on Offensive
  Technologies (WOOT 19)}}. \bibinfo{publisher}{USENIX Association},
  \bibinfo{address}{Santa Clara, CA}.
\newblock
\urldef\tempurl%
\url{https://www.usenix.org/conference/woot19/presentation/mambretti}
\showURL{%
\tempurl}


\bibitem[Mbed(2019)]%
        {mbed}
\bibfield{author}{\bibinfo{person}{Arm Mbed}.} \bibinfo{year}{2019}\natexlab{}.
\newblock \showarticletitle{MbedTLS: An open source, portable, easy to use,
  readable and flexible SSL library}.
\newblock \bibinfo{journal}{\emph{ARM Holdings plc}} (\bibinfo{year}{2019}).
\newblock


\bibitem[McFarlin et~al\mbox{.}(2013)]%
        {mcfarlin2013discerning}
\bibfield{author}{\bibinfo{person}{Daniel~S McFarlin}, \bibinfo{person}{Charles
  Tucker}, {and} \bibinfo{person}{Craig Zilles}.}
  \bibinfo{year}{2013}\natexlab{}.
\newblock \showarticletitle{Discerning the dominant out-of-order performance
  advantage: Is it speculation or dynamism?}
\newblock \bibinfo{journal}{\emph{ACM International Conference on Architectural
  Support for Programming Languages and Operating Systems (ASPLOS)}}
  (\bibinfo{year}{2013}).
\newblock


\bibitem[Molnar et~al\mbox{.}(2006)]%
        {molnar2006program}
\bibfield{author}{\bibinfo{person}{David Molnar}, \bibinfo{person}{Matt
  Piotrowski}, \bibinfo{person}{David Schultz}, {and} \bibinfo{person}{David
  Wagner}.} \bibinfo{year}{2006}\natexlab{}.
\newblock \showarticletitle{The program counter security model: Automatic
  detection and removal of control-flow side channel attacks}. In
  \bibinfo{booktitle}{\emph{Information Security and Cryptology-ICISC 2005: 8th
  International Conference, Seoul, Korea, December 1-2, 2005, Revised Selected
  Papers 8}}. Springer, \bibinfo{pages}{156--168}.
\newblock


\bibitem[Omar and Khan(2020)]%
        {Omar2021ironhide}
\bibfield{author}{\bibinfo{person}{Hamza Omar} {and} \bibinfo{person}{Omer
  Khan}.} \bibinfo{year}{2020}\natexlab{}.
\newblock \showarticletitle{IRONHIDE: A Secure Multicore that Efficiently
  Mitigates Microarchitecture State Attacks for Interactive Applications}. In
  \bibinfo{booktitle}{\emph{2020 IEEE International Symposium on High
  Performance Computer Architecture (HPCA)}}. \bibinfo{pages}{111--122}.
\newblock
\urldef\tempurl%
\url{https://doi.org/10.1109/HPCA47549.2020.00019}
\showDOI{\tempurl}


\bibitem[OpenSSL Ladder RSA(2011)]%
        {opensslrsaml}
OpenSSL Ladder RSA \bibinfo{year}{2011}\natexlab{}.
\newblock \bibinfo{title}{OpenSSL 1.01 Montgomery-{L}adder {RSA}}.
\newblock
  \bibinfo{howpublished}{\url{https://github.com/openssl/openssl/blob/46ebd9e3bb623d3c15ef2203038956f3f7213620/crypto/ec/ec2_mult.c}}.
\newblock


\bibitem[Patil et~al\mbox{.}(2021)]%
        {patil2021elfies}
\bibfield{author}{\bibinfo{person}{Harish Patil}, \bibinfo{person}{Alexander
  Isaev}, \bibinfo{person}{Wim Heirman}, \bibinfo{person}{Alen Sabu},
  \bibinfo{person}{Ali Hajiabadi}, {and} \bibinfo{person}{Trevor~E Carlson}.}
  \bibinfo{year}{2021}\natexlab{}.
\newblock \showarticletitle{ELFies: Executable region checkpoints for
  performance analysis and simulation}. In \bibinfo{booktitle}{\emph{2021
  IEEE/ACM International Symposium on Code Generation and Optimization (CGO)}}.
  IEEE, \bibinfo{pages}{126--136}.
\newblock


\bibitem[Puddu et~al\mbox{.}(2021)]%
        {puddu2021frontal}
\bibfield{author}{\bibinfo{person}{Ivan Puddu}, \bibinfo{person}{Moritz
  Schneider}, \bibinfo{person}{Miro Haller}, {and} \bibinfo{person}{Srdjan
  {\v{C}}apkun}.} \bibinfo{year}{2021}\natexlab{}.
\newblock \showarticletitle{Frontal Attack: Leaking $\{$Control-Flow$\}$ in
  $\{$SGX$\}$ via the $\{$CPU$\}$ Frontend}. In \bibinfo{booktitle}{\emph{30th
  USENIX Security Symposium (USENIX Security 21)}}. \bibinfo{pages}{663--680}.
\newblock


\bibitem[Qureshi(2018)]%
        {qureshi2018ceaser}
\bibfield{author}{\bibinfo{person}{Moinuddin~K Qureshi}.}
  \bibinfo{year}{2018}\natexlab{}.
\newblock \showarticletitle{CEASER: Mitigating conflict-based cache attacks via
  encrypted-address and remapping}. In \bibinfo{booktitle}{\emph{2018 51st
  Annual IEEE/ACM International Symposium on Microarchitecture (MICRO)}}. IEEE,
  \bibinfo{pages}{775--787}.
\newblock


\bibitem[Rane et~al\mbox{.}(2015)]%
        {rane2015raccoon}
\bibfield{author}{\bibinfo{person}{Ashay Rane}, \bibinfo{person}{Calvin Lin},
  {and} \bibinfo{person}{Mohit Tiwari}.} \bibinfo{year}{2015}\natexlab{}.
\newblock \showarticletitle{Raccoon: Closing digital side-channels through
  obfuscated execution}. In \bibinfo{booktitle}{\emph{24th $\{$USENIX$\}$
  Security Symposium ($\{$USENIX$\}$ Security 15)}}. \bibinfo{pages}{431--446}.
\newblock


\bibitem[RSA LibreSSL(2023)]%
        {libressl}
RSA LibreSSL \bibinfo{year}{accessed 9 April 2023}\natexlab{}.
\newblock \bibinfo{title}{RSA LibreSSL 3.3.6}.
\newblock
  \bibinfo{howpublished}{\url{https://github.com/PowerShell/LibreSSL/blob/edfc7eb366aa242e8966332e6d98a16cd4264d6c/crypto/bn/bn_exp.c\#L481}}.
\newblock


\bibitem[Saileshwar and Qureshi(2019)]%
        {saileshwar2019cleanupspec}
\bibfield{author}{\bibinfo{person}{Gururaj Saileshwar} {and}
  \bibinfo{person}{Moinuddin~K Qureshi}.} \bibinfo{year}{2019}\natexlab{}.
\newblock \showarticletitle{Cleanupspec: An" undo" approach to safe
  speculation}. In \bibinfo{booktitle}{\emph{Proceedings of the 52nd Annual
  IEEE/ACM International Symposium on Microarchitecture}}.
  \bibinfo{pages}{73--86}.
\newblock


\bibitem[Sakalis et~al\mbox{.}(2019)]%
        {DOM}
\bibfield{author}{\bibinfo{person}{Christos Sakalis}, \bibinfo{person}{Stefanos
  Kaxiras}, \bibinfo{person}{Alberto Ros}, \bibinfo{person}{Alexandra
  Jimborean}, {and} \bibinfo{person}{Magnus Sj{\"a}lander}.}
  \bibinfo{year}{2019}\natexlab{}.
\newblock \showarticletitle{Efficient invisible speculative execution through
  selective delay and value prediction}. In \bibinfo{booktitle}{\emph{2019
  ACM/IEEE 46th Annual International Symposium on Computer Architecture
  (ISCA)}}. IEEE, \bibinfo{pages}{723--735}.
\newblock


\bibitem[Schwarz et~al\mbox{.}(2020)]%
        {schwarz2020context}
\bibfield{author}{\bibinfo{person}{Michael Schwarz}, \bibinfo{person}{Moritz
  Lipp}, \bibinfo{person}{Claudio Canella}, \bibinfo{person}{Robert Schilling},
  \bibinfo{person}{Florian Kargl}, {and} \bibinfo{person}{Daniel Gruss}.}
  \bibinfo{year}{2020}\natexlab{}.
\newblock \showarticletitle{ConTExT: A Generic Approach for Mitigating
  Spectre.}. In \bibinfo{booktitle}{\emph{NDSS}}.
\newblock


\bibitem[Schwarz et~al\mbox{.}(2019)]%
        {schwarz2019zombieload}
\bibfield{author}{\bibinfo{person}{Michael Schwarz}, \bibinfo{person}{Moritz
  Lipp}, \bibinfo{person}{Daniel Moghimi}, \bibinfo{person}{Jo Van~Bulck},
  \bibinfo{person}{Julian Stecklina}, \bibinfo{person}{Thomas Prescher}, {and}
  \bibinfo{person}{Daniel Gruss}.} \bibinfo{year}{2019}\natexlab{}.
\newblock \showarticletitle{ZombieLoad: Cross-privilege-boundary data
  sampling}. In \bibinfo{booktitle}{\emph{Proceedings of the 2019 ACM SIGSAC
  Conference on Computer and Communications Security}}.
  \bibinfo{pages}{753--768}.
\newblock


\bibitem[Schwarz et~al\mbox{.}(2018)]%
        {NetSpectre2018Schwarz}
\bibfield{author}{\bibinfo{person}{Michael Schwarz}, \bibinfo{person}{Martin
  Schwarzl}, \bibinfo{person}{Moritz Lipp}, {and} \bibinfo{person}{Daniel
  Gruss}.} \bibinfo{year}{2018}\natexlab{}.
\newblock \showarticletitle{NetSpectre: Read Arbitrary Memory over Network}.
\newblock \bibinfo{journal}{\emph{CoRR}}  \bibinfo{volume}{abs/1807.10535}
  (\bibinfo{year}{2018}).
\newblock
\showeprint[arXiv]{1807.10535}
\urldef\tempurl%
\url{http://arxiv.org/abs/1807.10535}
\showURL{%
\tempurl}


\bibitem[Schwarzl et~al\mbox{.}(2021)]%
        {schwarzl2021specfuscator}
\bibfield{author}{\bibinfo{person}{Martin Schwarzl}, \bibinfo{person}{Claudio
  Canella}, \bibinfo{person}{Daniel Gruss}, {and} \bibinfo{person}{Michael
  Schwarz}.} \bibinfo{year}{2021}\natexlab{}.
\newblock \showarticletitle{Specfuscator: Evaluating Branch Removal as a
  Spectre Mitigation}.
\newblock \bibinfo{journal}{\emph{Financial Cryptography and Data Security
  2021}} (\bibinfo{year}{2021}).
\newblock


\bibitem[Van~Bulck et~al\mbox{.}(2018)]%
        {van2018foreshadow}
\bibfield{author}{\bibinfo{person}{Jo Van~Bulck}, \bibinfo{person}{Marina
  Minkin}, \bibinfo{person}{Ofir Weisse}, \bibinfo{person}{Daniel Genkin},
  \bibinfo{person}{Baris Kasikci}, \bibinfo{person}{Frank Piessens},
  \bibinfo{person}{Mark Silberstein}, \bibinfo{person}{Thomas~F Wenisch},
  \bibinfo{person}{Yuval Yarom}, {and} \bibinfo{person}{Raoul Strackx}.}
  \bibinfo{year}{2018}\natexlab{}.
\newblock \showarticletitle{Foreshadow: Extracting the keys to the Intel SGX
  kingdom with transient out-of-order execution}. In
  \bibinfo{booktitle}{\emph{Proceedings fo the 27th USENIX Security
  Symposium}}. USENIX Association.
\newblock


\bibitem[Van~Schaik et~al\mbox{.}(2019)]%
        {van2019ridl}
\bibfield{author}{\bibinfo{person}{Stephan Van~Schaik}, \bibinfo{person}{Alyssa
  Milburn}, \bibinfo{person}{Sebastian {\"O}sterlund}, \bibinfo{person}{Pietro
  Frigo}, \bibinfo{person}{Giorgi Maisuradze}, \bibinfo{person}{Kaveh Razavi},
  \bibinfo{person}{Herbert Bos}, {and} \bibinfo{person}{Cristiano Giuffrida}.}
  \bibinfo{year}{2019}\natexlab{}.
\newblock \showarticletitle{RIDL: Rogue in-flight data load}. In
  \bibinfo{booktitle}{\emph{2019 IEEE Symposium on Security and Privacy (SP)}}.
  IEEE, \bibinfo{pages}{88--105}.
\newblock


\bibitem[Vicarte et~al\mbox{.}(2022)]%
        {vicarte2022augury}
\bibfield{author}{\bibinfo{person}{Jose Rodrigo~Sanchez Vicarte},
  \bibinfo{person}{Michael Flanders}, \bibinfo{person}{Riccardo Paccagnella},
  \bibinfo{person}{Grant Garrett-Grossman}, \bibinfo{person}{Adam Morrison},
  \bibinfo{person}{Christopher~W Fletcher}, {and} \bibinfo{person}{David
  Kohlbrenner}.} \bibinfo{year}{2022}\natexlab{}.
\newblock \showarticletitle{Augury: Using data memory-dependent prefetchers to
  leak data at rest}. In \bibinfo{booktitle}{\emph{2022 IEEE Symposium on
  Security and Privacy (SP)}}. IEEE, \bibinfo{pages}{1491--1505}.
\newblock


\bibitem[Weisse et~al\mbox{.}(2019)]%
        {weisse2019nda}
\bibfield{author}{\bibinfo{person}{Ofir Weisse}, \bibinfo{person}{Ian Neal},
  \bibinfo{person}{Kevin Loughlin}, \bibinfo{person}{Thomas~F Wenisch}, {and}
  \bibinfo{person}{Baris Kasikci}.} \bibinfo{year}{2019}\natexlab{}.
\newblock \showarticletitle{NDA: Preventing speculative execution attacks at
  their source}. In \bibinfo{booktitle}{\emph{Proceedings of the 52nd Annual
  IEEE/ACM International Symposium on Microarchitecture}}.
  \bibinfo{pages}{572--586}.
\newblock


\bibitem[Weisse et~al\mbox{.}(2018)]%
        {weisse2018foreshadow}
\bibfield{author}{\bibinfo{person}{Ofir Weisse}, \bibinfo{person}{Jo
  Van~Bulck}, \bibinfo{person}{Marina Minkin}, \bibinfo{person}{Daniel Genkin},
  \bibinfo{person}{Baris Kasikci}, \bibinfo{person}{Frank Piessens},
  \bibinfo{person}{Mark Silberstein}, \bibinfo{person}{Raoul Strackx},
  \bibinfo{person}{Thomas~F Wenisch}, {and} \bibinfo{person}{Yuval Yarom}.}
  \bibinfo{year}{2018}\natexlab{}.
\newblock \showarticletitle{Foreshadow-NG: Breaking the virtual memory
  abstraction with transient out-of-order execution}.
\newblock  (\bibinfo{year}{2018}).
\newblock


\bibitem[Wikner and Razavi(2022)]%
        {wikner2022retbleed}
\bibfield{author}{\bibinfo{person}{Johannes Wikner} {and}
  \bibinfo{person}{Kaveh Razavi}.} \bibinfo{year}{2022}\natexlab{}.
\newblock \showarticletitle{$\{$RETBLEED$\}$: Arbitrary Speculative Code
  Execution with Return Instructions}. In \bibinfo{booktitle}{\emph{31st USENIX
  Security Symposium (USENIX Security 22)}}. \bibinfo{pages}{3825--3842}.
\newblock


\bibitem[Wu et~al\mbox{.}(2018)]%
        {wu2018eliminating}
\bibfield{author}{\bibinfo{person}{Meng Wu}, \bibinfo{person}{Shengjian Guo},
  \bibinfo{person}{Patrick Schaumont}, {and} \bibinfo{person}{Chao Wang}.}
  \bibinfo{year}{2018}\natexlab{}.
\newblock \showarticletitle{Eliminating timing side-channel leaks using program
  repair}. In \bibinfo{booktitle}{\emph{Proceedings of the 27th ACM SIGSOFT
  International Symposium on Software Testing and Analysis}}.
  \bibinfo{pages}{15--26}.
\newblock


\bibitem[Yan et~al\mbox{.}(2018)]%
        {yan2018invisispec}
\bibfield{author}{\bibinfo{person}{Mengjia Yan}, \bibinfo{person}{Jiho Choi},
  \bibinfo{person}{Dimitrios Skarlatos}, \bibinfo{person}{Adam Morrison},
  \bibinfo{person}{Christopher Fletcher}, {and} \bibinfo{person}{Josep
  Torrellas}.} \bibinfo{year}{2018}\natexlab{}.
\newblock \showarticletitle{Invisispec: Making speculative execution invisible
  in the cache hierarchy}. In \bibinfo{booktitle}{\emph{2018 51st Annual
  IEEE/ACM International Symposium on Microarchitecture (MICRO)}}. IEEE,
  \bibinfo{pages}{428--441}.
\newblock


\bibitem[Yu et~al\mbox{.}(2019a)]%
        {yu2019data}
\bibfield{author}{\bibinfo{person}{Jiyong Yu}, \bibinfo{person}{Lucas Hsiung},
  \bibinfo{person}{Mohamad El'Hajj}, {and} \bibinfo{person}{Christopher~W
  Fletcher}.} \bibinfo{year}{2019}\natexlab{a}.
\newblock \showarticletitle{Data Oblivious ISA Extensions for Side
  Channel-Resistant and High Performance Computing}. In
  \bibinfo{booktitle}{\emph{The Network and Distributed System Security
  Symposium (NDSS)}}.
\newblock


\bibitem[Yu et~al\mbox{.}(2020)]%
        {SDOSTT2020Yu}
\bibfield{author}{\bibinfo{person}{Jiyong Yu}, \bibinfo{person}{Namrata
  Mantri}, \bibinfo{person}{Josep Torrellas}, \bibinfo{person}{Adam Morrison},
  {and} \bibinfo{person}{Christopher~W. Fletcher}.}
  \bibinfo{year}{2020}\natexlab{}.
\newblock \showarticletitle{Speculative Data-Oblivious Execution: Mobilizing
  Safe Prediction for Safe and Efficient Speculative Execution}. In
  \bibinfo{booktitle}{\emph{Proceedings of the ACM/IEEE 47th Annual
  International Symposium on Computer Architecture}} (Virtual Event)
  \emph{(\bibinfo{series}{ISCA '20})}. \bibinfo{publisher}{IEEE Press},
  \bibinfo{pages}{707–720}.
\newblock
\showISBNx{9781728146614}
\urldef\tempurl%
\url{https://doi.org/10.1109/ISCA45697.2020.00064}
\showDOI{\tempurl}


\bibitem[Yu et~al\mbox{.}(2019b)]%
        {yu2019speculative}
\bibfield{author}{\bibinfo{person}{Jiyong Yu}, \bibinfo{person}{Mengjia Yan},
  \bibinfo{person}{Artem Khyzha}, \bibinfo{person}{Adam Morrison},
  \bibinfo{person}{Josep Torrellas}, {and} \bibinfo{person}{Christopher~W
  Fletcher}.} \bibinfo{year}{2019}\natexlab{b}.
\newblock \showarticletitle{Speculative taint tracking (stt) a comprehensive
  protection for speculatively accessed data}. In
  \bibinfo{booktitle}{\emph{Proceedings of the 52nd Annual IEEE/ACM
  International Symposium on Microarchitecture}}. \bibinfo{pages}{954--968}.
\newblock


\bibitem[Zhao et~al\mbox{.}(2020)]%
        {zhao2020invarspec}
\bibfield{author}{\bibinfo{person}{Zirui~Neil Zhao}, \bibinfo{person}{Houxiang
  Ji}, \bibinfo{person}{Mengjia Yan}, \bibinfo{person}{Jiyong Yu},
  \bibinfo{person}{Christopher~W. Fletcher}, \bibinfo{person}{Adam Morrison},
  \bibinfo{person}{Darko Marinov}, {and} \bibinfo{person}{Josep Torrellas}.}
  \bibinfo{year}{2020}\natexlab{}.
\newblock \showarticletitle{Speculation Invariance (InvarSpec): Faster Safe
  Execution Through Program Analysis}. In \bibinfo{booktitle}{\emph{2020 53rd
  Annual IEEE/ACM International Symposium on Microarchitecture (MICRO)}}.
  \bibinfo{pages}{1138--1152}.
\newblock
\urldef\tempurl%
\url{https://doi.org/10.1109/MICRO50266.2020.00094}
\showDOI{\tempurl}


\end{thebibliography}

\end{document}